\documentclass[prx, sd,amsmath,amssymb, reprint, superscriptaddress,longbibliography]{revtex4-1}
\usepackage{amsmath}
\usepackage{amssymb}
\usepackage{bbold}
\DeclareMathOperator{\sgn}{sgn}
\usepackage{etoolbox}
\usepackage{breqn}
\usepackage{graphicx}
\usepackage{subfigure}
\usepackage{dcolumn}
\usepackage{bm}
\usepackage{float}
\usepackage{hyperref}
\usepackage{enumerate}

\makeatletter
\patchcmd\eq@setnumber{\stepcounter}{\refstepcounter}{}{%
  \errmessage{Patching \noexpand\eq@setnumber failed}%
}
\makeatother
\makeatletter
\let\cat@comma@active\@empty
\makeatother
\let \oldnabla \nabla
\renewcommand{\nabla}{\bm{\oldnabla}}
\renewcommand{\vec}{\mathbf}

\newcommand{\vrho}{\bm \rho}
\newcommand{\hrho}{\hat{\bm\rho}}

\newcommand{\vom}{\bm \omega}
\newcommand{\haom}{\hat{\bm\omega}}
\newcommand{\om}{\omega}
\newcommand{\ve}{\bm \epsilon}

\newcommand{\vx}{\bm{\sigma}}
\newcommand{\vy}{\bm{\eta}}
\newcommand{\vz}{\bm{\nu}}

\newcommand{\vv}{\vec{v}}

\newcommand{\scF}{\mathcal{F}}
\newcommand{\scS}{\mathcal{S}}
\newcommand{\scO}{\mathcal{O}}

\newcommand{\hr}{\hat{\mathbf r}}
\newcommand{\uu}[1]{\mathbf{#1}}

\begin{document}
\title{Continuous versus Discontinuous Transitions in the $D$-Dimensional Generalized Kuramoto Model: Odd $D$ is Different}
\author{Sarthak Chandra}
\email{sarthakc@umd.edu}
\affiliation{Department of Physics, University of Maryland, College Park, MD, U.S.A.}
\author{Michelle Girvan}
\affiliation{Department of Physics, University of Maryland, College Park, MD, U.S.A.}
\affiliation{Institute for Physical Science and Technology, University of Maryland, College Park, MD, U.S.A.}
\author{Edward Ott}
\affiliation{Department of Physics, University of Maryland, College Park, MD, U.S.A.}
\affiliation{Department of Electrical and Computer Engineering, University of Maryland, College Park, MD, U.S.A.}
\begin{abstract}

The Kuramoto model, originally proposed to model the dynamics of many interacting oscillators, has been used and generalized for a wide range of applications involving the collective behavior of large heterogeneous groups of dynamical units whose states are characterized by a scalar angle variable. One such application in which we are interested is the alignment of orientation vectors among members of a swarm.
Despite being commonly used for this purpose, the Kuramoto model can only describe swarms in 2 dimensions, and hence the results obtained do not apply to the often relevant situation of swarms in 3 dimensions.
Partly based on this motivation, as well as on relevance to the classical, mean-field, zero-temperature Heisenberg model with quenched site disorder, in this paper we study the Kuramoto model generalized to $D$ dimensions.
We show that in the important case of 3 dimensions, as well as for any odd number of dimensions, the $D$-dimensional generalized Kuramoto model for heterogeneous units has dynamics that are remarkably different from the dynamics in 2 dimensions.
In particular, for odd $D$ the transition to coherence occurs discontinuously as the inter-unit coupling constant $K$ is increased through zero, as opposed to the $D=2$ case (and, as we show, also the case of even $D$) for which the transition to coherence occurs continuously as $K$ increases through a \emph{positive} critical value $K_c$. 
We also demonstrate the qualitative applicability of our results  to related models constructed specifically to capture swarming and flocking dynamics in three dimensions.

\end{abstract}
\maketitle
\section{Introduction} 
\subsection{Background}
Collective behavior in large populations of interacting elements has been a subject of intense study in physical, social, biological and technological systems\cite{Pikovsky2003, Montbrio2015,Vicsek2012,Carreras2004,Christakis2008,Wiesenfeld1998,Kiss2002,Abdulrehem2009,Motter2013,Chandra2017}. An important, frequently encountered example is the case of interacting phase oscillators, i.e., coupling between elements whose state is characterized by a point on a unit circle. In 1967 Winfree first systematically studied the dynamics of a population of weakly coupled phase oscillators\cite{Winfree1967}.
A few years later\cite{Kuramoto1975}, Kuramoto presented a simplified version of the Winfree model which he solved in the limit of $N\to\infty$, where $N$ is the number of oscillators. This model, now known as the Kuramoto model, is
\begin{equation}\label{eq:2dkuramoto}
\frac{d\theta_i}{dt} = \omega_i + \frac{K}{N}\sum_{j=1}^N \sin(\theta_j - \theta_i),
\end{equation}
where $\theta_i$ represents the phase angle of the $i^{\text{th}}$ oscillator, $\omega_i$ is its natural frequency of oscillation (which we will also refer to as the natural rotation), and $K$ is the coupling strength between oscillators. Typically the $\om_i$ are chosen randomly from some unimodal distribution with a finite spread $\Delta$, and $N\gg 1$ (the case of interest in this paper) is often considered.
In the $N\to\infty$ limit Kuramoto was able to show the presence of a continuous phase transition between asynchronous and partially synchronous states of the system\cite{Kuramoto1975,Kuramoto1984}. 

The Kuramoto model and its generalizations have since been used to study synchronization behavior in a wide variety of systems, modeling biological problems such as the behavior of cardiac pacemaker cells\cite{osaka2017}, synchronization in large groups of flashing fireflies\cite{Ermentrout1991, Buck1968}, circadian rhythms\cite{Antonsen2008, Childs2008}, and neuronal synchronization\cite{Acebron2005}, as well as problems in physics and engineering such as synchronization of power-grid networks\cite{Carreras2004, Motter2013}, superconducting Josephson junctions\cite{Marvel2009}, atomic physics\cite{Zhu2015}, and neutrino oscillation\cite{Pantaleone1998}, among others. Another class of applications of the Kuramoto model has been modeling the alignment of unit vectors representing the direction of motion of interacting members of a swarm or flock of moving agents in two dimensions\cite{o2017,Sepulchre2005,Chicoli2016}. Alternately, one can think of such unit vectors as characterizing the opinion of an individual in a group of interacting individuals\cite{OlfatiSaber2006}. In this later case, alignment of unit vectors can be viewed as modeling the evolution toward social consensus.

The aforementioned studies all describe the alignment of agents via a single scalar variable $\theta_i$, which characterizes the alignment state of the individual coupled units. However, for several of the above cited applications alignment in higher-dimensional spaces is important, and this is the subject of this paper. For example, the problem of alignment of velocity vectors in a flock of birds, a school of fish, or a swarm of flying drones is more realistically set up in three-dimensional space, whereas the alignment of opinion dynamics of a population could in general be multidimensional depending on the characteristics of the opinions considered. 
With such examples in mind, Olfati-Saber\cite{OlfatiSaber2006} introduced a higher-dimensional generalization of the Kuramoto model without the presence of any individual natural rotation [analogous to the $\om_i$ term in Eq. (\ref{eq:2dkuramoto})]. (In 2013, Zhu\cite{Zhu2013} considered the equivalent case of identical natural rotations for each agent.) 
The choice of the generalization in Refs. \cite{OlfatiSaber2006,Zhu2013} maintains the form of the coupling between two agents in all dimensions, i.e., in $D$ dimensions the state of each agent is taken to be a $D$-dimensional unit vector, and the coupling between two agents is proportional to the sine of the angle between their unit vectors\footnote{For generalizations of the Kuramoto model wherein agent states are represented by elements of a Lie group, see Refs.\cite{Lohe2009,Gu2007}.}.
Network characteristics leading to complete alignment were discussed; however, no complete stability analysis of the system was presented. In our paper we consider globally coupled systems, with a spread of the individual natural rotations of each unit, which follows from the generalization of the spread in the natural frequencies of the standard Kuramoto model.
These natural rotations act as constant biases to the states of the agents. In particular, for a given swarming agent, the natural rotation term can be thought of as a systematic error in the dynamics of the agent, which causes the agent to drift away from traveling in purely a straight line. We motivate the form natural rotation term in the context of flocking and swarming in $D=3$ in Sec. \ref{sec:fgdynamics}.
In assuming these natural rotations we set up a model more general than the one that has been studied by previous authors, leading to new and interesting results.

\subsection{Main Result}
\begin{figure}
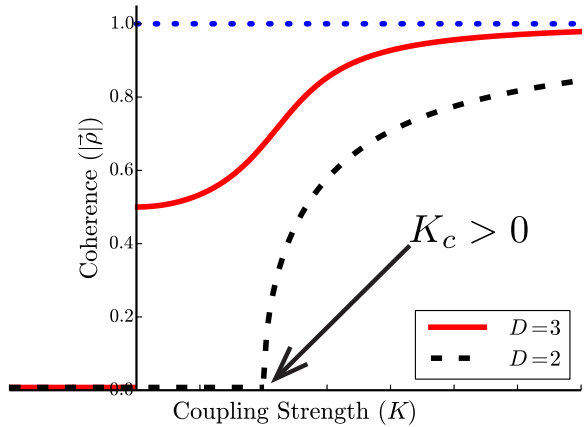

\includegraphics[width=\columnwidth]{{{Phase_transitions_in_2_and_3_dimensions_no_numbers_new}}}
\caption{Phase transitions for the standard two-dimensional Kuramoto model from theory (see Ref. \cite{Strogatz2000}), shown as the black dashed curve, and for the Kuramoto model generalized to three dimensions as calculated from the theory in Eq. (\ref{eq:rhoconsistency}), shown as the solid red curve. Note the continuous transition in the two-dimensional Kuramoto model at a critical coupling of $K_c>0$, and the discontinuous transition of the three-dimensional Kuramoto model at $K_c=0$. The blue dotted line represents the maximum possible value of coherence, corresponding to $|\vrho|=1$.}
\label{fig:phasetransition2and3}
\end{figure}

A key point in this paper is the remarkable difference between the standard two-dimensional Kuramoto model and its generalizations to 3 dimensions (and also to odd values of $D\geq 5$). A striking example of this is the nature of the transition from the incoherent state to the partially aligned state. As previously noted, the two-dimensional Kuramoto model, in the limit of infinite system size, was shown by Kuramoto\cite{Kuramoto1975,Kuramoto1984} to exhibit a continuous phase transition to coherence with increasing coupling strength $K$. This is represented by the dashed curve in Fig. \ref{fig:phasetransition2and3}, where the horizontal axis is the coupling strength, $K$, and the vertical axis represents the `order parameter' [Sec. \ref{sec:fullsetup}, defined in Eq. (\ref{eq:orderparameter})], which is a measure of the coherence (or degree of synchronization). The exact shape of this curve can be derived analytically\cite{Strogatz2000}, and it can be shown that this phase transition to synchrony is effectively a low-dimensional bifurcation\cite{Ott2008}.
The three-dimensional Kuramoto model, on the other hand, exhibits a \emph{discontinuous phase transition as we increase the coupling strength through zero} (solid curve in Fig. \ref{fig:phasetransition2and3}): For negative values of the coupling strength (indicative of repulsive interactions between agents), the agents tend to a completely incoherent state (defined by an `order parameter' with zero magnitude), and as we increase the coupling strength through zero, there is a discontinuous jump of the coherence as measured by the order parameter. Further, we find that this discontinuous phase transition occurs nonhysteretically.

\subsection{Relation to Statistical Physics Models}\label{sec:statphys}
It is interesting to note that if the time-independent frequencies $\om_i$ in Eq. (\ref{eq:2dkuramoto}) are replaced by independent, zero-mean, white noise of uniform strength, then the statistical equilibria and phase transitions of the Kuramoto model are the same as those of the mean-field classical XY model, which describes the interactions of classical two-dimensional spins with global coupling\cite{van1993,Uezu2015,Mehta2015,Collet2016}. In this case, the strength of the white noise corresponds to the temperature, and the magnitude of the coherence corresponds to the magnetization. Thus the Kuramoto model can be thought of as the mean-field XY model with thermal noise replaced by quenched randomness (the randomly chosen time-independent frequencies $\om_i$). Specifically, the mean-field XY model and the Kuramoto model yield similar behavior\cite{Uezu2015,Acebron2005} in that they both show a continuous (`second order') transition as the coupling constant increases through a critical value $K_c>0$ (which, for the Kuramoto model, increases with the spread $\Delta$ in the distribution of the natural frequencies, while, for the XY model, $K_c$ increases with temperature). A surprising result of our paper is that, when these models are extended to three dimensions, the two-dimensional qualitative similarity of the behavior for the cases of the quenched randomness and thermal noise versions of the XY model no longer applies: 	
  As mentioned above, the three-dimensional Kuramoto model with quenched disorder shows a discontinuous (`first order') phase transition at a zero coupling strength. Independent of the magnitude of the spread in the rotations comprising the quenched disorder, the three-dimensional Kuramoto model always shows partial alignment for $K>0$. Since in the three-dimensional model the coupling between any two agents is identical to the two-dimensional case, i.e., proportional to the sine of the angle between the unit vectors $\vx_i$ of the two agents, this model also describes the interactions of classical three-dimensional spins with global coupling, i.e., the mean-field classical Heisenberg model. If this quenched disorder in terms of the spread of natural rotations were to correspond with the a temporally noisy disordered system, then allowing for larger spread would correspond to higher temperatures and larger noise. 
However, at finite positive temperature the classical Heisenberg model, like the XY model, has a continuous phase transition at a positive critical coupling strength $K_c$\cite{Stanley1971}. Thus, in contrast to the two-dimensional case, for these problems in three dimensions there is a qualitative difference between temperature and quenched disorder.

\section{Model Description}\label{sec:fullsetup}

\begin{figure}
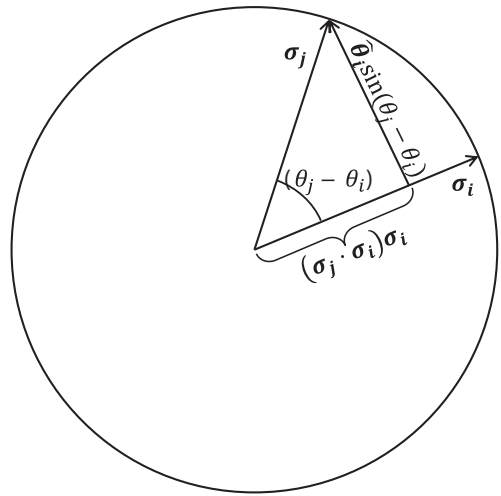

\includegraphics[width=0.75\columnwidth]{{{dot_product_illustration}}}
\caption{Illustration showing that $[\vx_j - (\vx_j\cdot\vx_i)\vx_i]=\hat{\bm\theta}_i \sin(\theta_j-\theta_i)$, where $\hat{\bm\theta}_i$ is a unit vector in the direction of increasing $\theta_i$.}
\label{fig:dotproduct}
\end{figure}

In order to see how the Kuramoto model can be generalized to higher dimensions\cite{OlfatiSaber2006, Zhu2013}, we note that Eq. (\ref{eq:2dkuramoto}) for $\theta_i$ can be rewritten (see Fig. \ref{fig:dotproduct} and its caption) in terms of the evolution of a collection of $N$ two-dimensional unit vectors, $\vx_i$ with $(x,y)$ components $(\cos\theta_i,\sin\theta_i)$:
\begin{equation}\label{eq:Ddim}
\frac{d\vx_i}{dt} = \frac{K}{N}\sum_{j=1}^N[\vx_j - (\vx_j\cdot\vx_i)\vx_i] + \uu{W}_i\vx_i,
\end{equation}
where 
\begin{equation}
\uu{W}_i= \begin{pmatrix} 0 & \omega_i \\ -\omega_i & 0\end{pmatrix}.
\end{equation}

From this point of view the natural generalization of the Kuramoto problem, Eq. (\ref{eq:2dkuramoto}), to $D$ dimensions is to consider Eq. (\ref{eq:Ddim}), but now with $\vx_i$ being a unit vector in $D$ dimensions and $\uu{W}_i$ being a real $D\times D$ antisymmetric matrix. Thus, unlike the standard Kuramoto model where the state of an agent is described by a single scalar variable $\theta_i$, the state of each agent in this generalized Kuramoto model is completely described by a $D$-dimensional unit vector $\vx_i$.

Each $\uu{W}_i$ term can be thought of as a constant bias to the dynamics of $\vx_i$. In the uncoupled dynamics, $d\vx_i/dt = \uu{W}_i\vx_i$, each agent is acted on by a constant linear operator, which causes each agent to move along the surface of the unit sphere $\scS$. For example, in the context of swarms or flocks, it is natural to assume that each agent, in the absence of coupling ($K=0$), has some imperfection that causes it to deviate away from the ideal of straight-line steady motion ($d\vx_i/dt = 0 $). Antisymmetry of $\uu{W}_i$ is imposed so as to ensure that the state vectors $\vx_i$ are unit vectors at all times. 

\begin{figure}
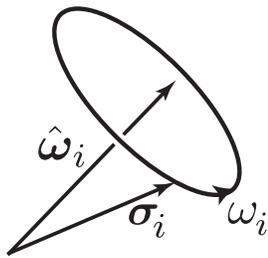

\includegraphics[width=0.4\columnwidth]{{{Precession_sigma}}}
\caption{$\vx_i$ precesses around $\haom_i$ with an angular frequency of $\om_i$ for $K=0$ with $D=3$.}
\label{fig:precession}
\end{figure}

For example, $D=3$, as discussed above, is of particular interest. In this case the term $\uu{W}_i\vx_i$ can be represented as
\begin{equation}
\uu{W}_i\vx_i=\vom_i\times\vx_i,\quad \vom_i=\om_i\haom_i, \; \om_i=|\vom_i|, \; \haom_i = \vom_i/|\vom_i|,
\end{equation}
where $\vom_i$ is referred to as the rotation vector; see Fig. \ref{fig:precession} which schematically represents the solution of Eq. (\ref{eq:Ddim}) for the case $K=0$ and $D=3$, in which $\vx_i$ is shown precessing around the vector $\haom_i=\vom_i/|\vom_i|$ at the rate $\om_i=|\vom_i|$. (Here, and later in this paper, we use the notation $|\mathbf{v}|$ to represent the Euclidean norm of the vector $\mathbf{v}$) Note that the dot product of the right-hand side of Eq. (\ref{eq:Ddim}) with $\vx_i$ is identically zero in all dimensions $D$, implying that $d|\vx_i|/dt=0$, consistent with $\vx_i$ being a unit vector.

In the context of the spin models discussed earlier in Sec. \ref{sec:statphys}, for positive $K$, the first term in Eq. (\ref{eq:Ddim}) corresponds to the interaction term between individual spins, with each pair of spins tending to align themselves parallel to each other. This term leads to macroscopic magnetization in the system of spins. The second term in Eq. (\ref{eq:Ddim}), $\uu{W}_i \vx_i$, corresponds to the quenched disorder discussed in Sec. \ref{sec:statphys} which inhibits coherence among the spins. 

In the context of flocking models, each $\vx_i$ is interpreted as the unit vector along the velocity vector for the $i^{\text{th}}$ agent. It is also assumed that the state of the agent is completely described by $\vx_i$, i.e., the agent is effectively axisymmetric about $\vx_i$. For positive $K$, the summation term in Eq. (\ref{eq:Ddim}) corresponds to all-to-all communication between agents in the flock, with each agent tending to align itself with each of the other agents. This term promotes coherence within the flock. The second term, $\uu{W}_i\vx_i$, corresponds to a simple dispersing term causing decorrelation of the agent orientations $\vx_i$. In particular, if we wish to consider the addition of a dispersal term that maintains the norm of $\vx_i$, and for simplicity is assumed to be time independent and linear, then it must be of the form $\uu{W}_i\vx_i$ for some antisymmetric matrix $\uu{W}_i$. 

In the context of swarms and flocks of three-dimensional agents, further motivation and justification for the form of the dispersing term $\uu{W}_i\vx_i$ is presented in Sec. \ref{sec:fgdynamics}. In particular, in order to support the possible generality of our main result (exemplified in Fig. \ref{fig:phasetransition2and3}), in Sec. \ref{sec:fgdynamics} we consider another model, different from the generalized Kuramoto model Eq. (\ref{eq:Ddim}), and show that our result also applies to this other model.

To better understand the dynamics of the generalized, $D$-dimensional Kuramoto model, we define an `order parameter', $\vrho$, that is analogous to the Kuramoto order parameter, $N^{-1}\sum_j \exp(i\theta_j)$, used to analyze the system of Eq. (\ref{eq:Ddim}) and is equivalent to it for $D=2$:
\begin{equation}\label{eq:orderparameter}
\vrho = \frac{1}{N}\sum_{i=1}^N \vx_i.
\end{equation}
Like the Kuramoto order parameter, $|\vrho|=1$ corresponds to the system being a completely coherent state, $\vx_i=\vx_j$ for all $i,j$; while $|\vrho|=0$ corresponds to an incoherent state. Using this order parameter, we can rewrite Eq. (\ref{eq:Ddim}) as
\begin{equation}\label{eq:Ddimrho}
\frac{d\vx_i}{dt} = K[\vrho - (\vrho\cdot\vx_i)\vx_i] + \uu{W}_i\vx_i.
\end{equation}
It can be seen in the above equation that the dynamics of each agent is determined by the two terms on the right-hand side. The first term (i.e., the term proportional to $K$) represents a global coupling of each agent to all the other agents through the order parameter. For positive $K$, this term attracts the state of each agent, $\vx_i$, towards the average orientation of the full population, characterized by the direction of $\vrho$; whereas for negative $K$, this term causes dispersal of the system agents away from coherence, with each agent moving away from the average orientation of the agents. The second term also gives dispersing dynamics, with each individual agent having distinct individual dynamics when uncoupled from the other agents in the system. 

To completely specify the setup of the system, we need to specify the choice of the $N$ natural rotations in Eq. (\ref{eq:Ddimrho}).
In the case of the standard $D=2$ Kuramoto model, Eq. (\ref{eq:2dkuramoto}), the natural rotations are added in the form of individual distinct natural frequencies $\om_i$ for each individual agent. Assuming that the natural frequency of each agent is independently picked randomly according to a fixed unimodal distribution $g(\omega)$, the change in coordinates, $\theta_i\to\theta_i+\om_0 t$, effectively reduces the natural frequency of each agent by any constant $\om_0$. Thus the mean of the distribution $g(\om)$ can be set to $0$ without loss of generality.
In the unit vector formulation of the $D=2$ Kuramoto model, this is equivalent to the change of variables $\vx_i\to e^{\uu{W_0} t}\vx_i$, where
\begin{equation*}
\uu{W_0} = \begin{pmatrix} 0 & \omega_0 \\ -\omega_0 & 0\end{pmatrix}.
\end{equation*} 
The new equation after the change of variables has the rotation matrix shifted as $\uu{W}_i \to e^{\uu{W_0} t} \uu{W}_i e^{-\uu{W_0} t} - \uu{W_0}$. In the case of $D=2$, the matrices $e^{\uu{W_0} t}$ and $\uu{W}_i$ commute, and hence the change is equivalent to the time-independent transformation $\uu{W}_i\to \uu{W}_i-\uu{W_0}$, allowing us to shift the mean of the distribution to any arbitrary quantity. For $D>2$, however, commutation of antisymmetric matrices or rotation matrices does not generally apply (i.e., the rotation group in $D>2$ is nonabelian), and hence this change of coordinates does not yield an equivalent model with a change of rotation matrices. Thus for $D>2$ the mean of this distribution cannot be simply shifted as in $D=2$.

In general, for $D$ dimensions, we specify the distribution over the space of antisymmetric matrices that we use to choose the individual $\uu{W}_i$ for each agent $i$. We denote this distribution by $G(\uu{W})$, which is analogous to the distribution $g(\omega)$ in the case of $D=2$. 
In this paper, we restrict the choice of $G(\uu{W})$ as follows: we choose each of the upper-triangular elements of $\uu{W}$ independently from a normal distribution with zero mean and a standard deviation of $\Delta$. The lower-triangular elements are then set according to the constraint that $\uu{W}$ is an antisymmetric matrix. This particular choice of $G(\uu{W})$ corresponds to an ensemble of antisymmetric matrices that has zero mean, and is invariant to rotations (choosing an anisotropic distribution, such as shifting the mean of the upper-triangular elements, or choosing the upper-triangular elements from normal distributions with unequal variance does not appear to change the qualitative results illustrated in Fig. \ref{fig:phasetransition2and3}). Hence, due to the rotational symmetry, $|\vrho|=0$ will be a solution to our system (note that this solution may be stable or unstable). 
Further, we also note that Eq. (\ref{eq:Ddim}) is invariant to the transformation $t\to\Delta\times t$, $K\to K/\Delta$ and $\uu{W}\to \uu{W}/\Delta$, and hence, without loss of generality, we set $\Delta$ to be unity for the remainder of this paper.

For future reference, it is useful to point out the following facts that apply to any real antisymmetric matrix $\uu{A}$ (such as $\uu{W}_i$):
\begin{enumerate}[(i)]
  \item Since $i\uu{A}$ is Hermitian, the real part of all the eigenvalues of $\uu{A}$ is zero. Hence all nonzero eigenvalues will be purely imaginary or zero.
	\item If $\lambda$ is an eigenvalue of $\uu{A}$, then so is $-\lambda$. \label{item2}
	\item If $D$ is odd, then $\uu{A}$ must have at least one zero eigenvalue [implied by (\ref{item2})]. Further, the corresponding eigenvector is real.
\end{enumerate}
For example, following Eq. (\ref{eq:Ddim}) we have noted that for $D=3$ we can express $\uu{W}_j\vx_j$ in the form $\vom_j\times\vx_j$, with $\vom_j=\om_j \haom_j$. In terms of the above discussion, $\haom_j$ is the real eigenvector corresponding to the zero eigenvalue of the $3\times3$ matrix $\uu{W}_j$ ($\uu{W}_j\haom_j=0$), and $\pm i \om_j$ are the nonzero eigenvalues of $\uu{W}_j$. 

As discussed earlier, for $D=3$ we can now represent the second term on the right-hand side of Eq. (\ref{eq:Ddimrho}) as a cross product, giving
\begin{equation}\label{eq:3dim}
\frac{d\vx_i}{dt} = K[\vrho - (\vrho\cdot\vx_i)\vx_i] + \vom_i\times \vx_i.
\end{equation}
Given the choice of the distribution $G(\uu{W})$ made above, we can write the distribution of the natural rotations of individual agents as $G(\vom)=g(\om)U(\haom)$, where $\vom=\om\haom$, with $\om=|\vom|$ and $\haom=\vom/\om$. The distribution of rotation directions, $U(\haom)$ is then isotropic, and independent of the distribution of rotation magnitudes, and the distribution of magnitudes is $g(\om)=\sqrt{2}\om^2\exp[-\om^2/(2\Delta^2)]/(\pi^{3/2}\Delta^3)$.	
This choice of the distribution $G(\vom)$ sets the mean of the distribution to always be $0$. In numerically simulating this system, we observe that the order parameter always goes to a fixed point, similar to the case of the standard Kuramoto model with zero mean of the distribution of frequencies. 

\section{Dynamics and Equilibria}\label{sec:dynamicsequilibria}

\begin{figure*}
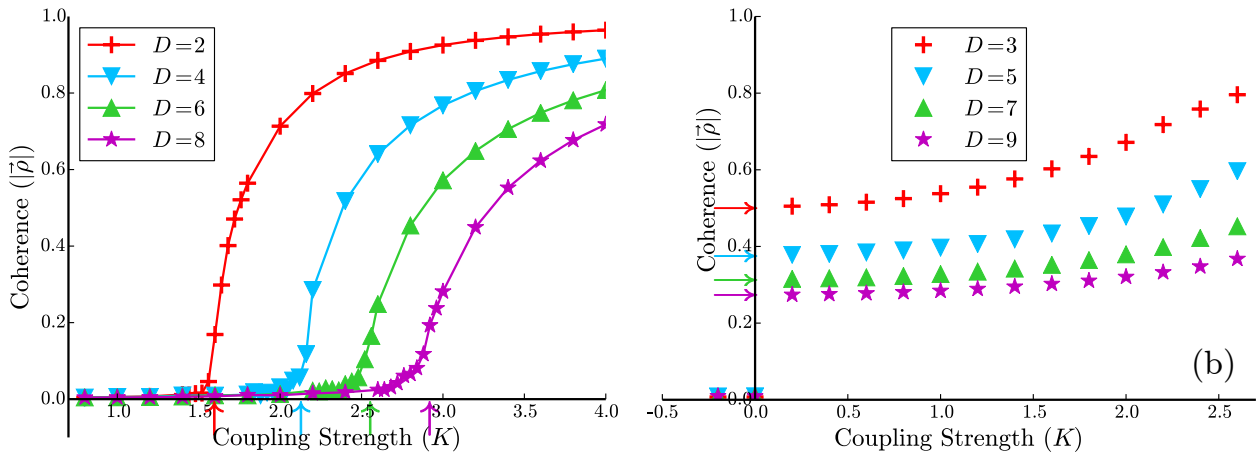

\includegraphics[width=2\columnwidth]{{{Phase_transitions_even_odd_w_Kc}}}
\caption{(a) Phase transitions for the generalized Kuramoto model for $D=2$ (red plus signs), $4$ (blue inverted triangles), $6$ (green triangles) and $8$	 (magenta stars) dimensions, numerical observations for $N=10^5$. (b) Phase transitions for the Kuramoto model generalized to $D=3$ (red plus signs), $5$ (blue inverted triangles), $7$ (green triangles) and $9$ (magenta stars) dimensions, numerical observations for $N=10^4$. $\Delta=1$ in each. 
The theoretical predictions from Eqs. (\ref{eq:kcgtilde}) and (\ref{eq:gtilde0}) for the critical coupling strength for even dimensions has been shown in correspondingly colored arrows on the $x$-axis in (a). For a discussion on the slight mismatch between the theory and the numerical results, see Sec. \ref{sec:evenD}. We expect this mismatch to decrease with increasing $N$.
The theoretical estimates from Eq. (\ref{eq:theorydiscontinuities}) for the magnitude of the discontinuity, i.e., $|\vrho|$ at $K\to0^+$ are shown in correspondingly colored arrows on the $y$-axis in (b). Note the close match between the theoretical result for the discontinuity, and the numerical observation for $|\vrho|$ at $K=0.2$. }
\label{fig:phasetransitionsoddeven}
\end{figure*}

To map out the interplay between the tendency to align and the natural rotation of the individual units [i.e. the two opposing tendencies represented by the two terms on the right-hand side of Eq. (\ref{eq:Ddim})], we plot numerically obtained phase transition diagrams for $D=2$ -- $9$ (see Fig. \ref{fig:phasetransitionsoddeven}). For large $N$ and varying values of the coupling strength $K$, we allow the system to reach its time asymptotic equilibrium, and then we plot the magnitude of the order parameter at equilibrium as a function of $K$. We note that the results in Fig. \ref{fig:phasetransitionsoddeven} apply for all the random initial realizations of the distributions of the individual states $\vx_i$ that we have tested.

As would be expected from the earlier discussion, in the case of negative coupling, i.e., $K<0$, the system of agents goes to a state which is incoherent, $|\vrho|\approx0$. 
For even $D$, as in the $D=2$ Kuramoto model (Fig. \ref{fig:phasetransitionsoddeven}), there exists a positive critical coupling constant $K_c>0$. In contrast, for odd $D$, coherence begins at $K=0$, i.e., $K_c=0$.
Moreover, in contrast to the even $D$ case where the transition is continuous (`second order'), for odd $D$ the transition is a discontinuous jump from $|\vrho|=0$ in $K<0$ to $|\vrho|>0$ for $K\to0^+$, past which $|\vrho|$ increases continuously with increasing $K$, asymptoting at $|\vrho|=1$ as $K\to \infty$. For example, for $D=3$ we find that $|\vrho|=0.5$ at $K=0^+$, and this result (as we shall subsequently show) is independent of the distribution $g(\om)$.
Furthermore, we find that this discontinuous transition is nonhysteretic. To better understand these observed phenomena, we now present a mathematical analysis of this system.

\subsection{Coherent states for $D=3$}\label{sec:coherentstates}

We first focus on the case of a positive coupling constant $K$ in three dimensions. We seek fixed points of the order parameter. To study these analytically we first solve for fixed points of the agents, assuming that the order parameter is at a fixed point with positive magnitude. We hence solve
\begin{equation}\label{eq:forvxF}
0=K[\vrho - (\vrho\cdot\vx_i^F)\vx_i^F] + \vom_i\times\vx_i^F
\end{equation}
for $\vx_i^F$.
The superscript $F$ indicates that the agent is at a fixed point. Given a spherically symmetric distribution of rotation vectors, we can choose the direction of the order parameter $\vrho$ arbitrarily. The magnitude of the order parameter must be chosen to be self consistent given the orientation of the agents, according to Eq. (\ref{eq:orderparameter}).
We define a quantity $\mu_i = \om_i/(K|\vrho|)$ to rewrite the above equation as
\begin{equation}\label{eq:forvxFscaled}
0=[\hrho - (\hrho\cdot\vx_i^F)\vx_i^F] + \mu_i (\haom_i\times\vx_i^F),
\end{equation}
where $\hrho=\vrho/|\vrho|$ is a unit vector in the direction of $\vrho$.
This vector equation can be solved (see Appendix \ref{apx:vxfproof}) to obtain
\begin{equation}\label{eq:rhodotx}
\hrho\cdot\vx_i^{F}=\pm \sqrt{\frac{(1-\mu_i^2) + \sqrt{(\mu_i^2-1)^2 + 4 \mu_i^2 (\hrho\cdot\haom_i)^2}}{2}},
\end{equation}
and in terms of $\hrho\cdot\vx_i^{F}$
\begin{equation}\label{eq:xexpression}
\vx_i^F = \frac{1}{1 + \xi_i^2 \mu_i^2} \left[ \mu_i(\haom_i \times\hrho) + \xi_i \mu_i^2 \haom_i + t_i\hrho\right]
\end{equation}
where $t_i=\hrho\cdot\vx_i^{F}$, and $\xi_i=\hrho\cdot\haom_i/\hrho\cdot\vx_i^{F}$.

From Eq. (\ref{eq:rhodotx}) we observe that there are two fixed points for each agent, one in the same hemisphere as the order parameter vector (corresponding to $\hrho\cdot\vx_i^{F} > 0$), and the other in the opposite hemisphere (corresponding to $\hrho\cdot\vx_i^{F} < 0$). Importantly, we also observe that there is a fixed point solution $\vx^F_i$ for any given $\vom_i$, $\vrho$ and $K$. Do these solutions correspond to a stable or unstable fixed points? Given a steady-state solution with all agents at one of their fixed points, for some $\vrho$ such that $|\vrho|>0$, we consider a perturbation $\ve_i$ to the $i^{\text{th}}$ agent. Assuming that $\vx_i(t)=\vx_i^F + \ve_i(t)$, we linearize Eq. (\ref{eq:3dim}) for small $\ve_i$ to obtain
\begin{equation}\label{eq:xstab}
\frac{d |\ve_i(t)|}{dt} = -K(\vrho\cdot\vx_i^F) |\ve_i(t)|,
\end{equation}
where we have used $\ve_i\cdot\vx_i^F=0$ so that the perturbed $\vx_i$ remains a unit vector.
Thus we see that the stability of the fixed point $\vx_i^F$ depends on the sign of $\vrho\cdot\vx_i^F$, with positive (negative) $\vrho\cdot\vx_i^F$ implying a stable (an unstable) fixed point. Since for each agent $\vx_i$ there are two solutions for $\vx_i^F$ with opposing signs of $\vrho\cdot\vx_i^F$ according to Eq. (\ref{eq:rhodotx}), each agent has a stable fixed point and an unstable fixed point. We assume that each agent will approach its stable fixed point.

\begin{figure*}
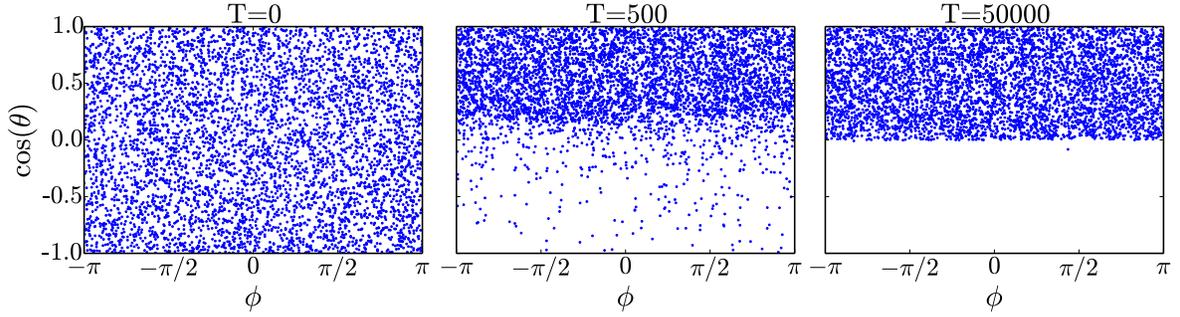

\includegraphics[width=1.8\columnwidth]{{{Rect_dists_new}}}
\caption{Orientations of each of the $N=5\cdot 10^3$ agents at time $T=0$, $T=500$ and $T=50000$. We visualize the orientations by plotting the endpoints of the orientation vectors on the sphere $\scS$. The sphere is then mapped onto a rectangle using an area-preserving transformation. We choose the $z$-axis along $\vrho$, and arbitrarily choose mutually orthogonal $x$ and $y$ axes. $\theta$ then represents the angle measured from the $z$-axis ($\cos\theta=\hrho\cdot\vx$), and $\phi$ represents the azimuthal angle measured anti-clockwise from the $x$-axis. The agent state vectors are initialized with a uniform distribution on the sphere, and evolved with a coupling constant $K=0.1$. Note how all of the agents tend to uniformly distribute themselves on one hemisphere.}
\label{fig:rectdists}
\end{figure*}

This behavior is in contrast to the two-dimensional Kuramoto model, where the proportion of agents in the entrained population increases continuously from $0$ as we increase $K$ beyond $K_c$. (This fundamental difference is due to the previously noted fact that $\uu{W}_i$ for odd $D$ always has zero as one of its eigenvalues.)
To understand the presence of the discontinuous phase transition, we first look at the case of small coupling, such that $0<K\ll \Delta$. By ignoring the first term on the right-hand side of Eq. (\ref{eq:forvxF}), or by considering the limit of $\mu_i\to\infty$ in Eqs. (\ref{eq:rhodotx}) and (\ref{eq:xexpression}), we see that $\vx_i^F=\pm\haom_i$. Since the stable fixed points corresponds to $\vrho\cdot\vx_i^F >0$, each agent will go to a stable fixed point given by $[\sgn(\vrho\cdot\haom_i)] \haom_i$. 
Note that this location of the fixed point on the unit sphere is independent of the magnitude of the agent's rotation vector, and depends only on the orientation of the rotation vector. Since the distribution of rotation vectors was chosen such that the distribution of directions $U(\haom)$ was uniform on the unit sphere, the fixed points $\sgn(\vrho\cdot\haom_i) \haom_i$ will be a set of uniformly distributed points over the hemisphere, $\vrho\cdot\vx>0$, of unit radius. This is demonstrated in Fig. \ref{fig:rectdists}, where we illustrate the orientations of $N=5\cdot10^3$ agents at a fixed time for a coupling strength $K=0.1$. In this plot, we have mapped the endpoints of the orientation vectors $\vx_i$ on the unit sphere $\scS$ to a rectangle via an area-preserving transformation (see Fig. \ref{fig:rectdists} caption for details). At the initial time, corresponding to an initial uniform distribution on $\scS$, the agents are uniformly distributed on the rectangle, whereas after $T=50000$ time units it can be seen that the agents are uniformly distributed over only the upper half of the rectangle, corresponding to the hemisphere $\vrho\cdot\vx>0$ of $\scS$.

As discussed earlier, the magnitude of the order parameter must be consistent with the orientations of the agents, according to Eq. (\ref{eq:orderparameter}). Thus, being the average of the orientations of all the agents, the order parameter will have $|\vrho|=1/2$, since the centroid of a hemisphere is located at a distance of half of the radius from the center of the sphere.
 
This result is independent of the choice of the distribution $g(\om)$, provided the rotation vector directions are isotropically distributed. As discussed earlier, negative values of coupling result in the system going to an incoherent state, with $|\vrho|=0$, while here we see that for small positive coupling the order parameter attains a value of $|\vrho|=0.5$. This result naturally generalizes to higher odd dimensions. As for the case of $D=3$, let $\haom_i$ be the real eigenvector corresponding to the zero eigenvalue of the $D\times D$ matrix $\uu{W}_i$. In the limit of $0<K\ll \Delta$, we can again ignore the first term on the right-hand of Eq. (\ref{eq:Ddim}). Solving for fixed points, we set $d\vx_i/dt=0$, and hence the fixed point solutions will be given by $\uu{W}_i\vx_i^F=0$, or $\vx_i^F=\pm\haom_i$. Following the same analysis as performed above for $D=3$, we reach the conclusion that for small positive $K$, the agents will go to fixed points given by $\sgn(\vrho\cdot\haom_i)\haom_i$. Hence the magnitude of $\vrho$ at $K=0^{+}$ will be given by the position of the centroid of a uniform hemisphere in $D$ dimensions:
\begin{equation}\label{eq:theorydiscontinuities}
|\vrho(K\to 0^+)| = \frac{2\Gamma(D/2)}{(D-1)\sqrt{\pi} \Gamma[(D-1)/2]}.
\end{equation}
This matches well with numerical results shown in Fig. \ref{fig:phasetransitionsoddeven} (b), where the colored arrows indicate the theory predictions according to Eq. (\ref{eq:theorydiscontinuities}). Note the close agreement with the prediction of the magnitude of $\vrho$ indicated by these arrows at $K=0$, and the $K>0$ start of the phase transition curves at $K=0.2$ shown by the various colored symbols.

\begin{figure}
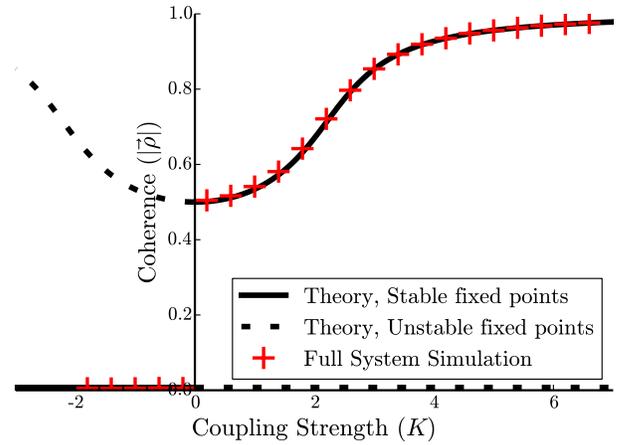

\includegraphics[width=\columnwidth]{{{phase_transition_theory_stable_unstable_scaled_new}}}
\caption{Phase transition for Kuramoto model generalized to three dimensions. The solid (dashed) black curve represents the derived stable (unstable) fixed points for the order parameter, and the red plus sign markers represent numerical results from a simulation with $N=10^4$ agents and $\Delta=1$. }
\label{fig:phasetransitiontheory}
\end{figure}

By setting up a consistency relation in a similar fashion (see below), we can calculate the magnitude of the order parameter as a function of the coupling constant for $D=3$. As shown in Fig. \ref{fig:phasetransitiontheory}, this theory (solid black curve) agrees well with results from simulations of Eq. (\ref{eq:Ddim}) with $N=10^4$ (red plus signs).

We now give our analysis for $D=3$ resulting in the solid curve in Fig. \ref{fig:phasetransitiontheory}. 
As earlier, we assume that $\vrho$ is in some particular fixed direction. Since the distribution of the direction of the unit vectors $\vx_i$ has been taken to be isotropic, we can assume that there will be rotational symmetry of the distribution of stable fixed points of the agents about the axis along $\vrho$. Thus
\begin{equation}\label{eq:sumconsistency}
|\vrho| = \frac{1}{N} \sum_{i=1}^N \hrho \cdot \vx_i^{F}.
\end{equation}
Since each agent has a unique natural rotation vector, we label the agent state variables as functions of their rotation vectors, as opposed to the index label $i$. Since the rotation vectors are chosen from a given distribution $G(\vom)$, we can approximate the above sum as 
\begin{equation}
|\vrho| = \int \hrho\cdot\vx^{F}(\vom) G(\vom) d\vom,
\end{equation}
which applies in the limit $N\to\infty$ in Eq. (\ref{eq:Ddim}).
We interpret $\hrho\cdot\vx^{F}(\vom)$ as $\cos[\theta(\vom)]$, where $\theta(\vom)$ is the angle between the direction of the order parameter, and the stable fixed point of the agent with rotation vector $\vom$. We will later use Eq. (\ref{eq:rhodotx}) to insert the expression for $\cos[\theta(\vom)]$. We write the above as
\begin{equation}
|\vrho| = \int \cos[\theta(\om_0,\haom)] g(\om_0) U(\haom) d\om_0 d\haom.
\end{equation}
Performing a change of variables from $\om_0$ to $\mu=\om/(K|\vrho|)$ we get
\begin{equation}
|\vrho| = \int \cos[\theta(\mu,\haom)] g(\mu K |\rho|) U(\haom) K |\vrho| d\mu d\haom,
\end{equation}
and hence
\begin{equation}\label{eq:rhoconsistency}
1 = \int \cos[\theta(\mu,\haom)] g(\mu K |\rho|) U(\haom) K d\mu d\haom.
\end{equation}
This can now be numerically solved to obtain $|\vrho|$ for a given $K$. For example, for the particular choice of $G(\vom)$ discussed above, where the three components of each vector $\vom_i$ are chosen independently from a normal distribution centered at $0$ with a standard deviation of $\Delta$, the integral in Eq. (\ref{eq:rhoconsistency}) over $\haom$ can be split into an azimuthal integral about the axis $\hrho$, which is trivial, and an integral over the angle between $\hrho$ and $\haom$, i.e., the $\zeta$ integral below
\begin{dmath}\label{eq:recursiveintegral}
1 = \frac{K}{2} \int_{-1}^{1} d\zeta \int_{-\infty}^{\infty} d\mu \sqrt{\frac{[1-\mu^2 + \sqrt{(\mu^2-1)^2 + 4 \mu^2 \zeta ^2}]}{2}} \times \frac{e^{-(\mu K |\vrho|)^2/(2\Delta^2)}}{(\sqrt{2\pi}\Delta)^3} (\mu K |\vrho|)^2 (2\pi), 
\end{dmath}
where the integration variable $\zeta$ represents $\hrho\cdot\haom$. Solving this integral equation numerically for $|\vrho|$ for different values of the coupling constant $K$ we obtain the solid black curve in Fig. \ref{fig:phasetransitiontheory}.

To complete the analysis of the coherent states for $K>0$, we now discuss why the state vectors $\vx_i$ approach their stable fixed points $\vx_i^F$. We demonstrate this in the limit of $0<K\ll \Delta$.
Under this assumption, we note that in Eq. (\ref{eq:3dim}), the typical magnitude of the second term on the right-hand side, $\mathcal{O}(\Delta)$, is much larger than the first term, which is $\mathcal{O}(|K\vrho|)$. We refer to $\mathcal{O}(\Delta)$ as the fast time-scale, and $\mathcal{O}(|K\vrho|)$ as the slow time-scale. 
The assumed separation of time-scales implies that, to lowest order, we can neglect the first term in Eq. (\ref{eq:3dim}), leading to the equation
\begin{equation}
\frac{d\vx_i}{dt} = \om_i \haom_i \times \vx_i.
\end{equation}
This has the solution depicted in Fig. \ref{fig:precession}, where the state vector $\vx_i$ uniformly precesses rapidly about $\haom_i$, with the quantity $z_i(t)=\vx_i(t)\cdot\haom_i$ constant on the fast time-scale. To determine the dynamics over the slow time-scale, we consider the dot product of Eq. (\ref{eq:3dim}) with $\haom_i$, and average both sides of the equation over the fast time scale. This gives the evolution of $z_i$ as 
\begin{equation}
\frac{dz_i(t)}{dt} = K \langle \vrho \rangle\cdot\haom_i [1-z_i(t)^2],
\end{equation}
where $\langle\vrho\rangle=N^{-1}\sum z_i\haom_i$, with the angle brackets representing averaging over the fast time-scale. This equation has a single stable fixed point at $+1$ or $-1$ dependent on the sign of $\langle \vrho \rangle\cdot\haom$. Thus starting from random initial conditions, $z_i(t)$ will move to its fixed point at $\sgn (\langle \vrho \rangle\cdot\haom)$. This is equivalent to stating that each $\vx_i$ will move to its fixed point $[\sgn(\vrho\cdot\haom_i)]\haom_i$. While we have only thus proved that $\vx_i$ will approach $\vx_i^F$ in the limit of small $K$, we numerically observe this to be true for all $K$, i.e., each agent goes to its corresponding stable fixed point as discussed above.

Until now, we have restricted our discussion of coherent states to $K>0$. Are there any stable coherent states for $K<0$? If the answer were yes, the fixed points of the agents could be calculated as earlier resulting in Eqs. (\ref{eq:rhodotx}) and (\ref{eq:xexpression}), and would be governed by the stability equation given in Eq. (\ref{eq:xstab}). Since $K<0$, the stable fixed points will correspond to solutions where $\vrho\cdot\vx^F < 0$ for each of the agents. This would imply that all of the agents would point to the hemisphere that the vector $\vrho$ points away from (not toward), contradicting the definition of $\vrho$ as the average of the orientations of all the agents. Thus there cannot be any stable fixed point solutions with positive magnitude of the order parameter for negative coupling. 

This, however, does not rule out the possibility of unstable coherent states with $K<0$. Going back to Eq. (\ref{eq:3dim}), we make a few observations. First, since all natural rotations were chosen such that the distribution of rotation directions was uniform on the sphere, the transformation $\vom\to-\vom$ does not affect the distribution or the macroscopic dynamics of the agents. After this transformation, we note that transforming $K\to-K$, and changing the direction of time, i.e., $t\to-t$, leaves Eq. (\ref{eq:3dim}) invariant. Thus, each stable fixed point of the macroscopic order parameter, $\vrho$, for a given value of coupling strength $K>0$, is also a fixed point at a coupling strength of $-K$, but is unstable (since we have reversed the sign of time). Thus the curve of coherent stable states for $K>0$ extends symmetrically to $K<0$ representing coherent unstable states. These stable (solid black curve) and unstable states (dashed black curve) are shown in Fig. \ref{fig:phasetransitiontheory}. We call these coherent states the `upper' branch of the phase transition diagram.

\subsection{Incoherent states for $D=3$}\label{sec:incoherentstates}

When the order parameter has zero magnitude, the system is said to be incoherent. As we demonstrate, this state is stable for negative values of the coupling constant and unstable for positive vales of the coupling constant.

In order to address the incoherent state, we first consider the following question: Given a state where $|\vrho|=0$ for all time, what are the possible dynamics of the individual agents? Setting $\vrho=0$ in Eq. (\ref{eq:Ddim}), we get $d\vx_i/dt = \uu{W}_i\vx_i$. In the case $D=3$, this means that the state $\vx_i$ of each agent precesses about their own rotation axes, as illustrated in Fig. \ref{fig:precession}. If each agent were randomly placed uniformly on $\scS$, then this would be consistent with $|\vrho|=0$, and would be a steady state. However, this is not the only such arrangement of $\vx_i$ that is possible corresponding to $|\vrho|=0$.
For example, if each agent, $\vx_i$ was placed on the axis of the corresponding rotation vector, such that $\vx_i=\haom_i$ (or $\vx_i=-\haom_i$), then this would also be consistent with $|\vrho|=0$ [since we have assumed that $U(\haom)$ is uniform], and the agents would each be at fixed points (this will be possible whenever $D$ is odd). In fact, the steady-state $|\vrho|=0$ applies for any random proportion $p$ of agents oriented parallel to the axes of their natural rotations, and the remaining agents at uniformly distributed locations on the sphere. Thus for $N\to\infty$ there are an infinite number of distributions of $\vom$ and $\vx$ for which $|\vrho|=0$ is a steady state.

To characterize these states in the limit of $N\to\infty$, we assume that the distribution of agent orientations $\vx$ rendered onto the unit sphere $\scS$ is well defined. We denote by $F(\vx,\vom,t)$ the distribution of agents on $\scS$, such that $F(\vx,\vom,t)d\vx d\vom$ is the fraction of agents that lie in the two-dimensional differential element on the surface $\scS$ centered at $\vx$ at time $t$, and have a natural rotation vector within the differential element $d\vom$ centered at $\vom$. Since the natural rotations of each agent are time independent and are independent of $\vx$, we can write
\begin{equation}\label{eq:fdefinition}
F(\vx,\vom,t)=G(\vom)f(\vx,\vom,t),
\end{equation}
where
\begin{equation*}
G(\vom) = \int_{\scS} F(\vx,\vom,t) d\vx
\end{equation*}
is the distribution of the antisymmetric natural rotation vectors, $\int G(\vom)d\vom=1$, and $G(\vom)=g(\om)U(\haom)$. In terms of this distribution function $F$, the order parameter will be given as
\begin{equation}\label{eq:ordparcts}
\vrho(t)=\int_{\scS} \vx G(\vom)f(\vx,\vom,t) d\vx d\vom.
\end{equation}
An example of a class of distributions in $D=3$ for which $|\vrho|=0$ is a steady state is given by 
\begin{dmath}\label{eq:classofdist}
F_0(\vx,\vom)=g(\om)U(\haom)\left[\frac{p}{2} \delta(\vx-\haom) + \frac{p}{2} \delta(\vx+\haom) + \frac{1-p}{4\pi}\right]\\=g(\om)U(\haom)f_0(\vx,\vom),
\end{dmath}
for any $p\in [0,1]$, where $\delta(\cdot)$ represents the Dirac delta function. 

As we will demonstrate shortly, in the limit $N\to\infty$, this entire class of distributions is stable to small perturbations for all $K<0$, i.e., for the incoherent region demonstrated in Fig. \ref{fig:phasetransition2and3}. This is in sharp contrast to the case of $D=2$, wherein there is a single stable incoherent steady-state distribution in the large system size limit (corresponding to $f=1/(2\pi)$) for the incoherent region in Fig. \ref{fig:phasetransition2and3}.

However, we observe from numerical simulations with $K<0$ (done at large, but necessarily finite $N$) that, starting with an initial condition corresponding to Eq. (\ref{eq:classofdist}) with $p=0$ (i.e., with $\vx_i$ distributed isotropically and \emph{independently} of its corresponding $\vom_i$, for all $i$) we observe that $\vx_i$ evolves slowly with time to either $\vx_i=+\haom_i$ or $\vx_i=-\haom_i$ (i.e., $\vx_i$ aligns with its rotation vector), with about half of the population $\{\vx_i\}$ going to $+\haom_i$, and half to $-\haom_i$. Furthermore, as $N$ increases, the rate of this relaxation becomes slower and slower, approaching zero as $N\to\infty$. In addition, the fractions of agents going to $+\haom_i$ and $-\haom_i$ approach $1/2$ as $N\to\infty$. 
Thus, taking the limit $t\to\infty$ followed by taking the limit $N\to\infty$, Eq. (\ref{eq:classofdist}) with $p=1$ (i.e., $F(\vx,\vom)=g(\om)U(\haom)[\delta(\vx-\haom)+\delta(\vx+\haom)]/2$) appears to approximate the distribution of agents on $\scS$. If the order in which the limits are taken is reversed, then $p=0$, its initial value (i.e., $F(\vx,\vom)=g(\om)U(\haom)U(\vx)$) represents the distribution of agents on $\scS$.
Similar results apply for other odd values of the dimension $D$, where $\haom_i$ is now the $D$-dimensional eigenvector of $\uu{W}_i$ having zero eigenvalue and with magnitude one (i.e., $\uu{W}_i\haom_i=0$). 

We illustrate these numerical results in Figs. \ref{fig:zplots} where we show the histograms of the initial (plotted in blue) and final (plotted in red) distributions of $\haom_i\cdot\vx_i$ over the $N$ agents. These numerical simulations were performed with $N=1000$, $K=-2$, $\Delta=1$. In the insets we plot the time-series of $z_i$ vs time for $50$ randomly chosen agents. We see that for all odd $D$, $\haom_i\cdot\vx_i$ evolves towards $\pm1$. Note that a similar consideration of even $D$ is inapplicable since a randomly chosen even-dimensional $\uu{W}_i$ typically does not have a zero eigenvalue, and thus $\haom_i$ does not exist.

\begin{figure*}
\includegraphics[width=2\columnwidth]{{{zplots_combined_scaled}}}
\caption{$N=1000$ agents were simulated with a coupling strength of $K=-2$. Histograms of $z_i=\vx_i\cdot\haom_i$ have been plotted at $T=0$ (in blue, corresponding to the initial condition having $\vx_i$ uniformly spread on $\scS$), and after $T=1.25\cdot10^5$ time units (in red). Note how the distributions concentrate at $1$ and $-1$ for large $T$. In the insets, we show plots of $z_i$ as a function of time for $50$ randomly chosen agents.}
\label{fig:zplots}
\end{figure*}

While macroscopically, in terms of the magnitude of the order parameter ($|\vrho|=0$), the $N\to\infty$ stationary states with distributions given by Eq. (\ref{eq:classofdist}) appear identical for all $p$, their stability to perturbations depends on $p$. To analyze the stability of this class of $N\to\infty$ stationary states we perform a linear analysis.
To do this, we first describe the dynamics of the system in terms of the distribution $F$. We treat Eq. (\ref{eq:3dim}) as a velocity field for the flow of this distribution and hence set up a continuity equation:
\begin{equation}\label{eq:Ddimcontinuity2}
\partial f/\partial t + \nabla_{\scS} \cdot (f(\vx,\vom,t) \vv) = 0,
\end{equation}
with a velocity field $\vv$ given by
\begin{equation}\label{eq:3dimvelocity}
\vv = K[\vrho - (\vx\cdot\vrho)\vx] + \vom\times\vx,
\end{equation}
where $\nabla_{\scS}\cdot\vec A$ represents the operator for the divergence of an arbitrary vector field $\vec A$, along the surface $\scS$ of the unit sphere in $\vx$-space.
The order parameter, $\vrho$ is described in terms of the distribution function $F$ according to Eq. (\ref{eq:ordparcts}).
We show in Appendix \ref{apx:continuity} that the continuity equation Eq. (\ref{eq:Ddimcontinuity2}) can be rewritten as 
\begin{dmath}\label{eq:distcontinuity}
\partial f/\partial t + [\nabla_{\scS} f(\vx,\vom,t) - 2f(\vx,\vom,t)\vx]\cdot \vrho  \\+ (\vom\times\vx)\cdot\nabla_{\scS} f(\vx,\vom,t)= 0,
\end{dmath}
where $\nabla_{\scS} \Phi$ is the component of the gradient of a scalar field $\Phi$ that is parallel to the surface $\scS$.
We consider a small perturbation, such that the distribution $f(\vx,\vom,t)$ can be written as
\begin{equation}\label{eq:perturbation}
f(\vx,\vom,t)=f_0(\vx,\vom) + \xi(\vx,\vom)e^{st},
\end{equation}
where $\xi(\vx,\vom)$ is small. Inserting Eq. (\ref{eq:perturbation}) into Eq. (\ref{eq:distcontinuity}) and linearizing gives 
\begin{equation}
s\xi(\vx,\vom,t) + (\vom\times\vx)\cdot\nabla \xi(\vx,\vom,t)= 2K(\vrho\cdot\vx)f_0(\vx,\vom).
\end{equation} 
To further simplify this equation, we make a choice of basis, such that $\vom=\om\hat z$. This allows us to rewrite the above equation as 
\begin{equation}\label{eq:perteqn}
s\xi(\vx,\vom,t) + \om \frac{\partial}{\partial \phi} \xi(\vx,\om,t)= 2K(\vrho\cdot\vx)f_0(\vx,\om),
\end{equation} 
where $\phi$ is the azimuthal coordinate around the $z$-axis. In this basis, we can then write $f_0$ as 
\begin{equation}\label{eq:classofdistspherical}
f_0(\theta,\phi,\vom)=\frac{p}{2} \frac{\delta(\theta) + \delta(\theta-\pi)}{\pi\sin(\theta)} + \frac{1-p}{4\pi},
\end{equation}
where $\theta$ is the angle measured from the $z$-axis, and together $\theta$ and $\phi$ represent $\vx$.

Inserting the form $f_0$ from Eq. (\ref{eq:classofdistspherical}) into Eq. (\ref{eq:perteqn}), we then solve for $\xi(\vx,\vom,t)$ and insert the obtained solution into Eq. (\ref{eq:ordparcts})  to obtain
\begin{equation*}
\vrho = \vrho (1-p) \frac{2K}{3}\left(\frac{1}{3s} + \frac{2s}{3}\int\frac{g(\om)d\om}{s^2 + \om^2}\right) + \vrho p \frac{2K}{3s},
\end{equation*}
giving the final dispersion relation,
\begin{equation}\label{eq:uniformdispersion}
1 =  (1-p) \frac{2K}{3}\left[\frac{1}{3s} + \frac{2s}{3}\int\frac{g(\om)d\om}{s^2 + \om^2}\right] + p \frac{2K}{3s}.
\end{equation}
Note that the case of $p=0$ (corresponding to an initial condition with independently chosen, uniformly random $\vx$) and the case of $p=1$ (corresponding to an initial condition with each $\vx$ being either $\haom$ or $-\haom$) have different dispersion relations. Thus, despite having the same macroscopic characteristic of $|\vrho|=0$, they will have different stabilities to perturbation. In the limit of small $K$, $s$ will also be small, and we can ignore the second term in the square brackets in the above expression. Thus,
\begin{equation*}
s=(1-p)\frac{2K}{9} + p \frac{2K}{3}.
\end{equation*}
Note that since $K$ is small, this represents the behavior of $s$ for $K$ around zero. Since $s\propto K$, we see that the incoherent state, having $|\vrho|=0$, will be stable ($s<0$) for $K<0$, and unstable ($s>0$) for $K>0$ as has been represented in Fig. \ref{fig:phasetransitiontheory}. We call these incoherent states the `lower' branch of the phase transition diagram.

It can be seen from Fig. \ref{fig:phasetransitiontheory} that the upper branch is stable whenever the lower branch is unstable (i.e., for $K>0$), and the upper branch is unstable whenever the lower branch is stable (i.e., for $K<0$). Thus, for no value of $K$ are there two values of $|\vrho|$ that are stable. This lack of bistability implies that the transition from incoherence to partial coherence occurs \emph{nonhysteretically} at $K=0$.

\subsection{Phase transition in even dimensions}\label{sec:evenD}

So far our primary focus has been on the cases of odd dimensions. As discussed earlier in Sec. \ref{sec:dynamicsequilibria} and in Fig. \ref{fig:phasetransitionsoddeven}, the even-dimensional cases exhibit continuous (`second order') phase transitions at positive critical coupling strength $K_c>0$. To better understand this, we extend the treatment of the $D=2$ case (e.g., Ref. \cite{Strogatz2000}) to even $D>2$. Like in the case of $D=3$, we assume that the system has reached an equilibrium, with the order parameter having a magnitude $|\vrho|$. Unlike Eq. (\ref{eq:xexpression}), wherein a fixed point for each agent exists for all values of $\uu{W}$, this will no longer be the case for even $D$. Rather, only certain values of the natural rotation $\uu{W}$ will permit the existence of fixed points above a certain value of $K$. Similar to Ref.\cite{Strogatz2000}, we first determine the conditions on $\uu{W}$ that permit fixed points of the corresponding agents, and then use this to set up a consistency relation similar to Eq. (\ref{eq:sumconsistency}) to determine $K_c$. A key assumption in this approach is that for steady states with $|\vrho|>0$ with $N\to\infty$, only agents for which $\vx$ is at a fixed point contribute to the sum in Eq. (\ref{eq:orderparameter}), which we prove \textit{a posteriori} [see Eqs. (\ref{eq:Cdefinition}) and (\ref{eq:fnormalization}) and accompanying discussion].

As in Eq. (\ref{eq:forvxF}), we see that the fixed points of $\vx_i$ must satisfy
\begin{equation}\label{eq:forvxFW}
0=K[\vrho - (\vrho\cdot\vx^F)\vx^F] + \uu{W}\vx^F,
\end{equation}
where we have dropped the index $i$ for simplicity. Denoting the term $(\vrho\cdot\vx^F)$ as $\gamma$, we observe
\begin{equation}
\vx^F = (\gamma\mathbb{1} - \uu{W})^{-1}\vrho,
\end{equation}
where $\mathbb{1}$ denotes the $D$-dimensional identity matrix.
Since $|\vx^F|^2=(\vx^F)^T\vx^F = 1$,
\begin{align}
1 &= \vrho^T (\gamma\mathbb{1} + \uu{W}/K)^{-1} (\gamma\mathbb{1} - \uu{W}/K)^{-1}\vrho \\
  &= \vrho^T (\gamma^2\mathbb{1} - \uu{W}^2/K^2)^{-1} \vrho. \label{eq:Hgammaoriginalbasis}
\end{align}
We now transform the above equation to a basis that block-diagonalizes the antisymmetric matrix $\uu{W}$. There exists a real orthogonal matrix, $\uu{R}$ such that $\uu{R}^T\uu{W}\uu{R}$ is a block-diagonal matrix whose $j^{\text{th}}$ block is the $2\times2$ matrix
\begin{equation*}
\uu{W^{(j)}} = \begin{pmatrix} 0 & \omega_j \\ -\omega_j & 0\end{pmatrix}
\end{equation*} 
for all $j\in\{1,2,\hdots,D/2\}$.
We will refer to these $\om_j$ as the $\Lambda=D/2$ natural frequencies associated with $\uu{W}$. Further, we define $\rho_k^2$ to be the sum of the squares of the magnitudes of the $2k-1^{\text{th}}$ and $2k^{\text{th}}$ components of $\uu{R}\vrho$.  Then Eq. (\ref{eq:Hgammaoriginalbasis}) can be simplified to
\begin{equation}\label{eq:Hgamma}
1 = \sum_{k=1}^{\Lambda} \frac{\rho_k^2}{\gamma^2 + \om_k^2/K^2} \equiv H(\gamma).
\end{equation}
Note that this change of basis does not affect the value of $\gamma$, since it is a scalar quantity.
Each term in the summand of the above expression can be interpreted as being proportional to a Lorentzian function of $\gamma$ centered about $\gamma=0$, and hence has a single maximum at $\gamma=0$. Thus, $H(\gamma)$ will also have a single maximum at $\gamma=0$, from which it follows that in order for Eq. (\ref{eq:Hgamma}) to have a real solution for $\gamma$, $H(\gamma=0)$ must be greater than or equal to $1$. Hence the condition on $\uu{W}$ that will permit the existence of $\vx^F$ will be 
\begin{equation}
H(\gamma=0)=K^2\sum_k{\frac{\rho_k^2}{\om_k^2}} > 1.
\end{equation}
For the case of the standard $D=2$ Kuramoto model, the above criteria reduces to $|\omega|<|K\vrho|$ (Ref. \cite{Strogatz2000}, Eq. (4.2)). For a given $\vrho$, we denote the region in $\uu{W}$-space that satisfies that the above criteria as $\Gamma$. Each $\uu{W}_i\in\Gamma$ will have a corresponding fixed point for $\vx_i$ and the set of such agents $i$ will be referred to as the entrained population. For each $\uu{W_j}\notin\Gamma$, $\vx_j$ is continually in motion, and we refer to these agents as the drifting population. We now argue that the contribution to the order parameter, $\vrho$ from the drifting population will be zero, and then use the Eq. (\ref{eq:Hgamma}) to write out a consistency relation for the order parameter as calculated only from the remaining entrained population.

Assuming an equilibrium of the system, such that the order parameter is at a fixed point, the drifting agents must form a stationary distribution on $\scS$. We denote this distribution by $f(\vx,\uu{W})$, which is analogous to $f(\vx,\vom,t)$ defined in Eq. (\ref{eq:fdefinition}). Since the velocity of each agent is governed by Eq. (\ref{eq:Ddimrho}), stationarity of the distribution requires that $f(\vx,\uu{W})$ is inversely proportional to the magnitude of this velocity. Hence
\begin{equation}\label{eq:Cdefinition}
f(\vx,\uu{W}) = \frac{C(\uu{W},K\vrho)}{|K[\vrho - (\vrho\cdot\vx)\vx] +\uu{W}\vx |},
\end{equation}
where $C(\uu{W},K\vrho)$ is a normalization constant,
\begin{equation}\label{eq:fnormalization}
\int_{|\vx|=1} f(\vx,\uu{W}) d\vx= 1
\end{equation}
for each $\uu{W}$ not in $\Gamma$. Since $\Gamma$ is invariant to the transformation $\uu{W}\to-\uu{W}$, it follows from the definition of $C(\uu{W},K\vrho)$ that it must also be invariant to $\uu{W}\to-\uu{W}$. The contribution to the order parameter from the drifting population is then given by 
\begin{equation*}
\vrho_{\text{drift}} = \int_{|\vx|=1} \int_{\uu{W}\notin\Gamma} \vx \frac{C(\uu{W},K\vrho)}{|K[\vrho - (\vrho\cdot\vx)\vx] +\uu{W}\vx |} G(\uu{W}) d\uu{W} d\vx.
\end{equation*}
Applying the variable transformations of $\vx\to-\vx$ and $\uu{W}\to-\uu{W}$ we obtain $\vrho_{\text{drift}}=-\vrho_{\text{drift}}$, and hence $|\vrho_{\text{drift}}|=0$.

Thus, the only contribution to the order parameter is from the entrained population of agents.
Let $H(\gamma)=1$ give rise to some solution $(\vrho\cdot\vx^F)=\gamma\equiv\gamma(\{\om_i\},\{\rho_i\})$. Then, dotting both sides of Eq. (\ref{eq:orderparameter}) with $\vrho$ in the limit of infinite system size gives
\begin{equation}\label{eq:rhosqoriginal}
|\vrho|^2 = \int_{\Gamma} \gamma(\{\om_i\},\{\rho_i\}) G(\uu{W})d\uu{W}. 
\end{equation}
As in the two-dimensional case, the critical coupling strength, $K_c$, will be such that the magnitude of the order parameter is infinitesimally small but nonzero. We can use this to determine a value of the critical coupling as 
\begin{equation}\label{eq:kcgtilde}
K_c = \frac{2}{\pi \tilde g(0)},
\end{equation}
where 
\begin{equation}
\tilde g(0) = \int_{-\infty}^\infty \hdots \int_{-\infty}^\infty   g(0,\om_2,\hdots,\om_{\Lambda})  d\om_2\hdots d\om_\Lambda,
\end{equation}
and $g(\om_1,\hdots,\om_{\Lambda})$ is the joint distribution of natural frequencies associated with the distribution $\uu{W}$ (see Appendix \ref{apx:evenDproof} for details). Note that, for our particular choice of an antisymmetric matrix ensemble from which we randomly draw the $\uu{W}_i$ (i.e., independently Gaussian upper-triangular matrix elements), there are known results for $g$ and $\tilde g$ from random matrix theory. In particular, Ref. \cite{Mehta1968} yields\footnote{In comparing the equation in Ref. \cite{Mehta1968}, with our numerical results, we observe that there appears to be a misprint of a factor of $1/(D\sqrt{2})$ in their expression} 
\begin{equation}\label{eq:gtilde0}
\tilde g(0) = \frac{1}{D}\sqrt{\frac{2}{\pi}} \sum_{n=0}^{(D/2)-1} \frac{(2n)!}{2^{2n} (n!)^2}.
\end{equation}
The predictions for the critical coupling strength, $K_c$, made according to Eqs. (\ref{eq:kcgtilde}) and (\ref{eq:gtilde0}) for $D=2, 4, 6$ and $8$ have been marked by vertical arrows in Fig. \ref{fig:phasetransitionsoddeven}(a). We expect that with increasing $N$ the numerically observed transitions will appear to be sharper at the marked critical coupling strength. 
Note that continuing the curve from large values of $|\rho|$ to the $x$-axis without changing its curvature (as would be expected from the shape of the phase transition curve in $D=2$; see Fig. \ref{fig:phasetransition2and3}) approximates the predicted values accurately.

\section{Model variant: Extended-body agents in three dimensions}\label{sec:fgdynamics}

\begin{figure}
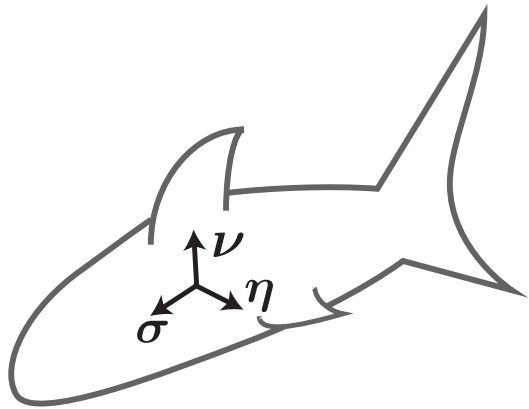

\includegraphics[width=0.8\columnwidth]{{{fish_with_axes_nu}}}
\caption{Illustration of an extended-body agent. Unlike the agents in the generalized Kuramoto model Eq. (\ref{eq:Ddim}), we assume that the state of an extended-body agent cannot be described by a single unit vector $\vx$. Rather, the pair of vectors $\vx$ and $\vy$ together describe the orientation and state of the agent. The direction of agent velocity is assumed to be along the direction $\vx$ as earlier. The unit vector $\vz$ is defined as $\vz = \vx\times \vy$}
\label{fig:rigidagent}
\end{figure}

From Eq. (\ref{eq:Ddim}), the dynamics of the system of agents can be thought of resulting from the interplay of two terms, $K[\vrho - (\vrho\cdot\vx_i)\vx_i]$, promoting coherence among agents, and $\uu{W}_i\vx_i$, promoting decoherence between agents. We have shown that the competition between these two opposing tendencies is resolved by a critical transition from incoherence to coherence that is qualitatively different for even and odd dimensionality (Figs. \ref{fig:phasetransition2and3} and \ref{fig:phasetransitionsoddeven}). In order to show that this qualitative result is not restricted to our particular assumed form of the $K=0$ agent dynamics ($d\vx_i/dt = \uu{W}_i\vx_i$), we here consider a very different model with $D=3$, and show that our conclusion for the behavior shown for the solution of Eq. (\ref{eq:Ddim}) continues to apply. 
Specifically, we consider a different form of the dispersal term in the context of the three-dimensional dynamics of extended objects (e.g., the fish in Fig. \ref{fig:rigidagent}). We will also further justify the term $\uu{W}_i\vx_i$ as a simple choice of dispersive dynamics for interacting agents.

As we discuss in Sec. \ref{sec:fullsetup}, the setup of Eq. (\ref{eq:Ddim}) considers interactions between agents that are fully described by a single $D$-dimensional unit vector. For an extended object, a single unit vector does not uniquely specify the agent state. In the specific context of three-dimensional extended objects in three-dimensional space (e.g., the dynamics of flocks of birds, swarms of drones etc.), the orientation of the extended body must be specified by two unit vectors. We call such agents extended-body agents, and describe their state via the two vectors $\vx$, which as earlier represents the direction of the velocity of the extended-body agent; and $\vy$, chosen orthogonal to $\vx$ (see Fig. \ref{fig:rigidagent}). For simplicity, we define $\vz=\vx\times\vy$ to form the right-handed orthonormal triple $\{\vx,\vy,\vz\}$. 
It should be noted that extended-body agents in two dimensions are completely described by a single unit vector, $\vx$, as in the standard $D=2$ Kuramoto model.
We will first set up the dynamics of this extended-body agent when it is not coupled to other agents. Motivated by the uncoupled dynamics of this extended-body agent representing some fixed errors/biases, we assume that the uncoupled dynamics of this extended-body agent is autonomous, i.e., not explicitly dependent on time. Under this assumption, we write
\begin{align}
d\vx/dt &= \Phi(\vx,\vy), \nonumber\\
d\vy/dt &= \Psi(\vx,\vy), \label{eq:PhiPsiTheta}\\
d\vz/dt &= \Theta(\vx,\vy) = \Phi \times \vy + \vx\times\Psi \nonumber.
\end{align}
Further, we make the natural assumption that the dynamics do not depend on information of its orientation with respect to any fixed frame of reference, i.e., there is no `special' direction in space that determines the dynamics of the extended-body agent. Thus,

\begin{align}
\Phi(\uu{R}\vx,\uu{R}\vy) &= \uu{R}\Phi(\vx,\vy), \label{eq:Phirotinvar} \\
\Psi(\uu{R}\vx,\uu{R}\vy) &= \uu{R}\Psi(\vx,\vy), \\
\Theta(\uu{R}\vx,\uu{R}\vy) &= \uu{R}\Theta(\vx,\vy), 
\end{align}
for any rotation matrix $\uu{R}$. 
Since the unit vectors $\{\vx,\vy,\vz\}$ form an orthonormal basis, we can write the vector field $\Phi$ in this basis,
\begin{equation}
\Phi(\vx,\vy) = a(\vx,\vy)\vx + b(\vx,\vy)\vy + c(\vx,\vy)\vz.
\end{equation}
Using Eq. (\ref{eq:Phirotinvar}), 
\begin{align}
a(\uu{R}\vx,\uu{R}\vy)\uu{R}\vx + b(\uu{R}\vx,\uu{R}\vy)\uu{R}\vy + c(\uu{R}\vx,\uu{R}\vy)\uu{R}\vz \nonumber\\
= \uu{R}[a(\vx,\vy)\vx + b(\vx,\vy)\vy + c(\vx,\vy)\vz].
\end{align}
Comparing components along $\uu{R}\vx$ on both sides of the above equation, 
\begin{equation}
a(\uu{R}\vx,\uu{R}\vy) = a(\vx,\vy),
\end{equation}
and hence the scalar $a(\vx,\vy)$ must be independent of $\vx$ and $\vy$, $a(\vx,\vy)=a$. Similarly, $b(\vx,\vy)$ and $c(\vx,\vy)$ must also be independent of $\vx$ and $\vy$. Applying similar reasoning to all the components of $\Phi(\vx,\vy)$, $\Psi(\vx,\vy)$ and $\Theta(\vx,\vy)$, we see that they must each be linear functions of $\vx$, $\vy$ and $\vz$. Hence [noting that $\uu{R}\vz = (\uu{R}\vx)\times (\uu{R}\vy)$]
\begin{align}
\Phi(\vx,\vy) &= a \vx + b \vy + c \vz, \nonumber\\
\Psi(\vx,\vy) &= a' \vx + b' \vy + c' \vz, \label{eq:linearized}\\
\Theta(\vx,\vy) &= a'' \vx + b'' \vy + c'' \vz. \nonumber
\end{align}

Further, since $\{\vx,\vy,\vz\}$ are unit vectors forming a right-handed triple,
\begin{equation}\label{eq:stayunit}
\vx\cdot\Phi = \vy\cdot \Psi = \vz\cdot\Theta = 0, 
\end{equation}
and, using $d/dt(\vx\cdot\vy)=d/dt(\vx\cdot\vz)=d/dt(\vy\cdot\vz)=0$,
\begin{equation}\label{eq:stayperp}
\vx\cdot\Psi + \vy\cdot\Phi = \vx\cdot\Theta + \vz\cdot\Phi = \vy\cdot\Theta + \vz\cdot\Psi = 0.
\end{equation}
Thus, Eqs. (\ref{eq:linearized}) reduce to
\begin{align}
d\vx/dt &= \Phi(\vx,\vy)  =-\alpha \vy + \beta\vz ,\nonumber \\
d\vy/dt &= \Psi(\vx,\vy)  = \alpha \vx + \gamma\vz, \label{eq:uncoupledrigid}\\
d\vz/dt &= \Theta(\vx,\vy)=-\beta\vx   - \gamma\vy. \nonumber
\end{align}
for some scalar, extended-body-agent specific quantities $\alpha$, $\beta$ and $\gamma$. Having specified the uncoupled dynamics of an extended-body agent, we add the effect of inter-agent coupling, in the form of the Kuramoto-like interactions described in Sec. \ref{sec:fullsetup}. Thus, analogous to Eq. (\ref{eq:Ddim}), we write
\begin{equation}\label{eq:rigidx}
\frac{d\vx_i}{dt} = K[\vrho - (\vrho\cdot\vx_i)\vx_i] + \Phi_i(\vx_i,\vy_i),
\end{equation}
where $\vrho$ is given by Eq. (\ref{eq:orderparameter}), which is the average of the velocity directions $\vx_i$ of each extended-body agent. Note that this form of coupling treats $\vx$ as a special direction as compared to $\vy$ and $\vz$, since we assume that the goal of the swarm is to maintain coherence via coupling that aligns the velocity direction $\vx_i$ of each agent $i$ to the the motion of the swarm as a whole.
We then write $\dot{\vy}_i$ and $\dot{\vz}_i$ such that the constraint Eqs. (\ref{eq:stayunit}) and (\ref{eq:stayperp}) continue to hold for the coupled system and that $K=0$ corresponds to Eqs. (\ref{eq:PhiPsiTheta}).
\begin{align}
d\vy_i/dt &= -K[\vrho\cdot\vy_i]\vx_i + \Psi_i(\vx_i,\vy_i) , \label{eq:rigidy}\\
d\vz_i/dt &= -K[\vrho\cdot\vz_i]\vx_i + \Theta_i(\vx_i,\vy_i) . \label{eq:rigidz}
\end{align}

\begin{figure}
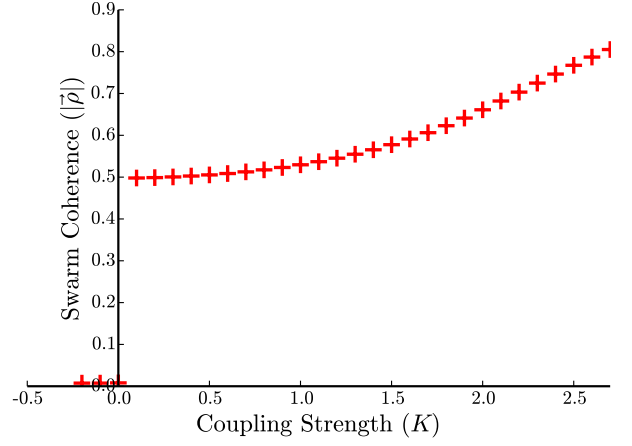

\includegraphics[width=\columnwidth]{{{f_g_Phase_transitions_in_3_dimensions}}}
\caption{Phase transition for interacting three-dimensional extended-body agents described by Eqs. (\ref{eq:rigidx}) -- (\ref{eq:rigidz}). The dynamics of individual agents in this system have been constructed to satisfy constraints imposed by extended-body dynamics, and are \emph{not} equivalent to the dynamics of the generalized Kuramoto model described in Sec.\ref{sec:dynamicsequilibria}. Despite this, we continue to observe a discontinuous jump in the asymptotic steady-state value of $|\vrho|$ as $K$ in increased through 0.}
\label{fig:fgtransition}
\end{figure}

We perform a simulation of $N=10^4$ such extended-body agents by numerically integrating Eqs. (\ref{eq:rigidx}), (\ref{eq:rigidy}) and (\ref{eq:rigidz}) for a range of values of $K$ similar to Fig. \ref{fig:phasetransitionsoddeven}. Since the dynamics captured by the Eqs. (\ref{eq:PhiPsiTheta}) represent random biases/errors, we choose the quantities $\alpha$, $\beta$ and $\gamma$ for each agent from independent, normal distributions with zero mean and unit variance. In Fig. \ref{fig:fgtransition} we present the phase transition displayed by this system of evolving extended-body agents. For each value of $K$ we numerically integrate the system until $|\vrho|$ reaches a steady-state value. Note that we continue to observe a discontinuous transition of $|\vrho|$ as $K$ increases through $0$. Further, we also numerically observed that if $\alpha$, $\beta$ and $\gamma$ are chosen anisotropically, i.e., if they are chosen from normal distributions with zero mean but differing variance, the qualitative result shown in Fig. \ref{fig:fgtransition} does not change, i.e., the transition to coherence is still discontinuous at $K=0$. This indicates that the phenomenon of discontinuous transitions in odd dimensions is not specific to the form of the dispersal term chosen in Eq. (\ref{eq:Ddim}), rather, it is a more general phenomena occurring for a potentially wide range of systems of interacting agents in odd dimensions. In contrast, this model for two dimensions ($\beta=\gamma=0$) is the same as the original Kuramoto model and hence has a continuous transition to coherence at a critical positive value of $K$.

To further examine the dynamics of the uncoupled agents Eqs. (\ref{eq:uncoupledrigid}), we adopt notation where we represent Eqs. (\ref{eq:uncoupledrigid}) as
\begin{equation}\label{eq:vxvyvzeqn}
\frac{d}{dt} \begin{pmatrix}\vx & \vy & \vz\end{pmatrix} = \begin{pmatrix}\vx& \vy& \vz\end{pmatrix}\uu{U} ,
\end{equation}
where $\uu{U}$ is the $3\times 3$ antisymmetric matrix,
\begin{equation}
\uu{U}=\begin{pmatrix}%
0 & \alpha & -\beta \\
-\alpha & 0 & -\gamma \\
\beta & \gamma & 0 \end{pmatrix},
\end{equation}
and $\begin{pmatrix}\vx & \vy & \vz\end{pmatrix}$ represents a $3\times 3$ matrix whose columns are the vectors $\vx$, $\vy$ and $\vz$.
We can then consider a change of basis $\uu{R}$ such that 
\begin{equation}
\uu{R}^T\uu{U}\uu{R}=\begin{pmatrix}%
0 & \om & 0 \\
-\om & 0 & 0 \\
0 & 0 & 0 \end{pmatrix},
\end{equation}
where $\om^2=\alpha^2 + \beta^2 + \gamma^2$. Using the same convention as Eq. (\ref{eq:vxvyvzeqn}), we define the orthonormal triple of unit vectors $\{\vec{u},\vec{v},\vec{w}\}$ as 
\begin{equation}\label{eq:uvwdef}
\begin{pmatrix}\vec{u}& \vec{v}& \vec{w}\end{pmatrix} = \begin{pmatrix}\vx& \vy& \vz\end{pmatrix}\uu{R}.
\end{equation}
Thus Eq. (\ref{eq:vxvyvzeqn}) becomes
\begin{align}
d\vec{u}/dt &=  \om \vec{v}, \nonumber \\
d\vec{v}/dt &= -\om \vec{u}, \label{eq:uvweqns}\\
d\vec{w}/dt &=0. \nonumber
\end{align}
According to Eqs. (\ref{eq:uvweqns}), the vectors $\vec{u}$ and $\vec{v}$, which are fixed linear combinations of $\vx$, $\vy$ and $\vz$, undergo uniform rotation with an angular frequency of $\om$ about the axis $\vec{w}$. Since $\vx$, $\vy$ and $\vz$ describe the physical orientation of the uncoupled extended-body agent, the extended-body agent will demonstrate dynamics that correspond to rotations in three dimensions, and there will exist a $\uu{W}$ such that $\dot\vx = \uu{W}\vx$. In particular, the axis of this rotation will be along the unit vector $\vec{w}$ given by $\vec{w}=R_{13}\vx + R_{23}\vy + R_{33}\vz$, where $R_{ij}$ is the $ij^{\text{th}}$ component of $\uu{R}$. Since $\dot{\vec{w}} = 0$, 
\begin{equation*}
\vec{w}=R_{13}\vx(0) + R_{23}\vy(0) + R_{33}\vz(0).
\end{equation*}
Note that $\uu{R}$ (and hence its components) is dependent on the random biases/systematic errors present, arising from the particular form of $\Phi$ and $\Psi$, whereas $\vx(0), \vy(0)$ and $\vz(0)=\vx(0)\times \vy(0)$ depends on the initial orientation/state of the extended-body agent. Thus the axis of rotation is dependent on the initial state of the extended-body agent, while the frequency of rotation, $\om$ is determined solely by the random systematic errors of the extended-body agent [i.e., $\alpha$, $\beta$ and $\gamma$ in Eq. (\ref{eq:uncoupledrigid})].

Thus, under the assumptions made above, the dynamics of uncoupled extended-body agents can be described as $\dot{\vx}=\uu{W}\vx$ for some initial-condition-dependent $\uu{W}$ (in particular, $\uu{W}\vx = -\om \vec{w}\times\vx$). Note however that this is $\emph{not}$ identical to the uncoupled dynamics of the agents described in Eq. (\ref{eq:Ddim}). In particular, the axis of rotation of the extended-body agent under the dynamics described here is along the vector $\vec{w}$, which is determined by the initial conditions of the extended-body agent state $(\vx(0),\vy(0))$. However, in the uncoupled dynamics of the generalized Kuramoto agents described by Eq. (\ref{eq:Ddim}) the axis of rotation is predetermined by the rotation matrix $\uu{W}_i$ assigned to agent $i$ and is independent of the initial condition chosen for the agent. An isotropic ensemble of rotation matrices for the generalized Kuramoto agents in the case of extended-body agents corresponds to an `isotropic' distribution of the extended-body agent parameters ($\alpha,\beta,\gamma$), as well as isotropic initial conditions of the extended-body agents. 

Further, this simple interpretation of $\vx$ undergoing uniform rotation no longer holds for the case of coupled extended-body agents, and Eqs. (\ref{eq:rigidx}), (\ref{eq:rigidy}) and (\ref{eq:rigidz}) cannot be simply written in the form of Eq. (\ref{eq:Ddim}) with an initial-condition-dependent $\uu{W}$ for arbitrary $K$ (this, however, is possible in the limit of $K\to 0$ or $|\vrho|\to 0$, hence our results for the stability analysis of the $|\vrho|=0$ state will recreate the phase transitions in higher dimensions). The qualitative dynamics of coupled extended-body agents and coupled generalized Kuramoto agents described by Eq. (\ref{eq:Ddim}) are in general distinct, yet our main point of discontinuous phase transitions for odd dimensions at $K=0$ continues to hold.

Thus, for the case of extended-body agents, under the assumptions made in this section, rotation matrices as dispersal terms arise naturally as simple error/fixed-bias terms for the individual agents. Rather than considering the case of initial-condition-dependent rotation matrices, in Eq. (\ref{eq:Ddim}) we have considered the simplification of choosing fixed rotation matrices $\uu{W}$. This motivates the generalization of the Kuramoto model presented in Eq. (\ref{eq:Ddim}) as a simple model to capture the dynamics of swarming and flocking agents. Further, we also see that the result obtained from our toy model Eq. (\ref{eq:Ddim}) for the qualitative continuous or discontinuous behavior of the incoherent-to-coherent transition continues to hold for other three-dimensional agent dynamics, such as the extended-body agent dynamics described in this section.

In Sec. \ref{sec:conclusions} we briefly describe other extensions and variants to the generalized Kuramoto model described in Sec. \ref{sec:fullsetup}.

\section{Discussion and conclusions}\label{sec:conclusions}

We have considered a generalization of the Kuramoto model to arbitrary dimensions, describing a system of interacting, orientable units, whose state is completely described by $D$-dimensional unit vectors. Our main result (Fig. \ref{fig:phasetransitionsoddeven}) is that the macroscopic dynamics of the Kuramoto model is strongly dependent on the dimensionality of the system, with odd-dimensional systems behaving similar to one other, and likewise for even-dimensional systems. For odd-dimensional systems, including the practically important case of $D=3$, we find that the phase transition from incoherence to partially coherent states occurs via a discontinuous, nonhysteretic transition as the coupling strength $K$ increases through 0 (Sec. \ref{sec:coherentstates}, also see Fig. \ref{fig:phasetransitiontheory}). In contrast, even-dimensional systems, like $D=2$, numerically appear to undergo continuous transitions of the coherence at a critical coupling strength $K_c>0$ (Fig. \ref{fig:phasetransitionsoddeven} (a)). We also note that, unlike the two-dimensional Kuramoto model, the state of the system is not always completely classified by the magnitude of the order parameter. In particular, for the two-dimensional Kuramoto model there is a single stable incoherent steady-state distribution in the infinite size limit ($f=1/(2\pi)$), whereas the three-dimensional Kuramoto model has an infinite number of such distributions (for example, Eq. (\ref{eq:classofdist})) each with different stability properties (see Eq. (\ref{eq:uniformdispersion})).
By considering a setup of extended-body agents, in Sec. \ref{sec:fgdynamics} we further motivated our choice of model Eq. (\ref{eq:Ddim}) in the context of swarms of drones or flocks of birds. In particular, we demonstrated that our qualitative results relating to the difference between odd- and even-dimensional systems continue to hold for models that use a different choice of the dispersal term. This study of extended-body agents in Sec. \ref{sec:fgdynamics} also explains why the choice of the dispersal term $\uu{W}\vx$ in the context of the qualitative phase transitions observed for $D=3$ is justified .

While other authors\cite{OlfatiSaber2006, Zhu2013} have also studied the Kuramoto model generalized to higher dimensions, their consideration has been limited to the case of identical natural rotations. Our setup of the problem (i.e., with heterogeneous natural rotations) by setting $G(\uu{W})=\delta(\uu{W}-\uu{W_0})$ reproduces the results in Refs. \cite{OlfatiSaber2006, Zhu2013} for the case of globally coupled systems (here we interpret the Dirac delta function acting on the antisymmetric matrix $\uu{W}$ as the product of Dirac delta functions acting on each of the upper-triangular elements of the matrix individually).  This heterogeneous setup of the problem now describes the interplay of two opposing tendencies, i.e., the tendency for agent states to align due to the inter-agent coupling, \emph{and} the tendency for agents to disperse themselves in opposition to such alignment. This leads to the possibility of new and interesting phenomena such as the difference between the odd and even dimensionality described in this paper.

In addition to the variant described in Sec. \ref{sec:fgdynamics}, the setup of the generalized Kuramoto model given by Eqs. (\ref{eq:Ddimrho}) and (\ref{eq:orderparameter}) can be modified and generalized in various ways. An interesting question for possible future study is whether a striking difference between odd and even dimensions (as we have found for the generalized Kuramoto model and its variant in Sec. \ref{sec:fgdynamics}) manifests in these modifications. 
For example, beyond the globally coupled systems we have considered, one might consider network-based coupling, wherein agent $j$ influences agent $i$ with a strength $A_{ij}$.
This is equivalent replacing $\vrho$ in Eq. (\ref{eq:Ddimrho}) with $\vrho_i$, where
\begin{equation*}
\vrho_i=\frac{1}{N} \sum_j A_{ij} \vx_j.
\end{equation*}
In the context of swarms of drones, a further natural generalization would be to have the network-based coupling $A_{ij}$ depend on the spatial distance and relative orientation between the $i^{\text{th}}$ and $j^{\text{th}}$ swarm agent. 

As discussed earlier, for positive $K$ the dynamics of each $\vx_i$ are attracted towards the average state of the system, $\vrho$. This could be interpreted as a target direction for each $\vx_i$, and can be generalized by replacing Eq. (\ref{eq:orderparameter}) by other definitions of $\vrho$. For example, in the context of swarms of drones, it could be desirable for the orientation of the drones to be biased towards the plane of the horizon, or to be biased toward the direction of a given target destination. To achieve this, the `target direction', $\vrho$ in Eq. (\ref{eq:Ddimrho}) could be modified from the average state of the system to the average state biased towards a given target.
Studying the dependence of the dynamics of such swarms of agents on modifications to $\vrho$ (via either the presence of network dependent interaction, or other bias targets) would be an interesting line of future research. 

In a future paper\cite{Chandra2018} we will present a mathematical formulation for studying the $D$-dimensional Kuramoto model in the infinite size limit via a generalization of the Ott-Antonsen ansatz\cite{Ott2008, Ott2009}, wherein we will also address the issue of generalization of $\vrho$.

\section*{Acknowledgements}

We thank Thomas M. Antonsen for useful discussion. We also thank the referees for their useful comments.
This work was supported by ONR grant N000141512134 and by AFOSR grant FA9550-15-1-0171.

\appendix

\section{Equation for fixed points of agents}\label{apx:vxfproof}
We here present a derivation of Eq. (\ref{eq:xexpression}). 
In what follows in this appendix, we write the fixed point solution of the $i^{\text{th}}$ agent, i.e., $\vx_i^{F}$ in Eq. (\ref{eq:forvxFscaled}) as simply $\vx$. We also similarly drop the index $i$ from $\mu_i$ and $\haom_i$ for simplicity of notation.

Taking the second term on the left-hand side of Eq. (\ref{eq:xexpression}) to the right-hand side and considering the square of the norm of both sides, we obtain
\begin{equation}\label{eq:1}
 [1 - (\hrho\cdot\vx)^2] = [1 - (\haom\cdot\vx)^2]\mu^2.
\end{equation}
Also, dotting Eq. (\ref{eq:forvxFscaled}) with $\haom$ we obtain
\begin{equation}\label{eq:2}
\hrho\cdot\haom = (\hrho\cdot\vx)(\haom\cdot\vx)
\end{equation}
Using Eq. (\ref{eq:2}) to replace the term $(\haom\cdot\vx)$ in Eq. (\ref{eq:1}) we obtain
\begin{equation}
1 -  (\hrho\cdot\vx)^2 = \left( 1 - \frac{(\hrho\cdot\haom)^2}{(\hrho\cdot\vx)^2} \right) \mu^2.
\end{equation}

Thus we have
\begin{equation}
1 - (\hat\rho\cdot\vx)^2 = \mu^2 - \frac{(\hat \rho\cdot\haom)^2}{(\hat\rho\cdot\vx)^2}\mu^2 = 0,
\end{equation}
which is a quadratic equation in $(\hat\rho\cdot\vx)^2$, whose solution is Eq. (\ref{eq:rhodotx}).
For $K>0$ the positive solution Eq. (\ref{eq:rhodotx}) will be stable, as is argued in the text. Equation (\ref{eq:forvxFscaled}) dotted with $\hrho$ gives
\begin{equation}
[1 - (\hrho \cdot \vx)^2] + \mu\hrho\cdot(\haom\times \vx) = 0.
\end{equation}
This can be rewritten using Eq. (\ref{eq:1}) as
\begin{equation}\label{eq:wlogbef}
\haom\cdot[\mu\haom - \mu(\haom\cdot\vx)\vx +  \vx\times\hrho ] = 0.
\end{equation}
Keeping $\haom$ fixed, we can independently choose $K$, and hence $\mu$. Thus the term in Eq. (\ref{eq:wlogbef}) in the square brackets must be independently zero.
\begin{equation}
\mu\haom - \mu(\haom\cdot\vx)\vx + \vx\times\hrho = 0.
\end{equation}
Using Eq. (\ref{eq:2}) again we obtain
\begin{equation}\label{eq:vrhocrossvx}
\hrho \times \vx = \mu\left( \haom - \xi \vx \right),
\end{equation}
where 
\begin{equation}\label{eq:xiexpression}
\xi = \frac{\hrho\cdot\haom}{\hrho\cdot\vx}.
\end{equation}

Since the solution to $\vec a \times \vec b = \vec c$, is $\vec b = (\vec c \times \vec a)/|\vec a|^2 + t \vec a$ for any $t$,

\begin{equation}\label{eq:3}
\vx = ( (\haom\times\hrho) - \xi(\vx\times\hrho) )\mu + t \hrho.
\end{equation}
Dotting both sides of Eq. (\ref{eq:3}) with $\hrho$, we see that $t=\hrho\cdot\vx$, which was solved for earlier, resulting in Eq. (\ref{eq:rhodotx}). We now go back to Eq. (\ref{eq:3}) and use Eq. (\ref{eq:vrhocrossvx}) to obtain
\begin{equation}
\vx = ( \mu(\haom\times\hrho) + \mu^2\xi(\haom - \xi \vx) ) + t \hrho,
\end{equation}
which can be rearranged to give
\begin{equation}
\vx = \frac{1}{1 + \xi^2 \mu^2} \left[ \mu(\haom \times\hrho) + \xi\mu^2 \haom + t\hrho\right],
\end{equation}
with $t=\hrho\cdot\vx$ and $\xi$ according to Eq. (\ref{eq:xiexpression}). This completes our derivation of Eq. (\ref{eq:xexpression})

\section{Simplification of continuity equation}\label{apx:continuity}

In this appendix we give a derivation of Eq. (\ref{eq:distcontinuity}) from Eq. (\ref{eq:Ddimcontinuity2}). We present this proof in arbitrary dimensions, where we rewrite Eq. (\ref{eq:3dimvelocity}) as
\begin{equation}\label{eq:Ddimvelocity}
\vv = K[\vrho - (\vx\cdot\vrho)\vx] + \uu{W}\vx,
\end{equation}
where $\vv$ is defined on the $(D-1)$-dimensional surface of the unit sphere $\scS$ embedded in $D$ dimensions. To simplify the continuity equation for the flow along the surface $\scS$, i.e., Eq. (\ref{eq:Ddimcontinuity2}), we first extend the velocity flow field to the entire space $\mathbb{R}^D$ by allowing $\vx$ to be a general $D$-vector (rather than restricting it to a unit vector). We then write the continuity equation using the regular divergence defined over the entire space, and demonstrate that this reduces to Eq. (\ref{eq:distcontinuity}) when considered on the surface $\scS$.

We write $\vx=r\hr$. Let the velocity flow field as extended to $\mathbb{R}^D$ be
\begin{align}
\vv_{\vx} &= \uu{W}\hr +  K[\vrho - \hr(\vrho \cdot \hr)], \\
        &= \uu{W}\vx/r +   K[\vrho - \vx (\vrho \cdot \vx)/r^2].
\end{align}
Note that this extension to $\mathbb{R}^D$ can be performed in multiple ways and does not affect our final result. Since $\hr\cdot \vv_{\vx} =0$, this flow field maintains the surfaces of spheres centered at $r=0$ as invariant manifolds. We then extend the distribution $f(\vx,t)$, that was defined on the surface $\scS$, to the entire space $\mathbb{R}^D$ as
\begin{equation}
\scF (\hr, r,\uu{W}, t) = f(\hr,\uu{W},t)\delta(r-1),
\end{equation}
where $\delta(\cdot)$ is the Dirac delta function. We can write the continuity equation for the flow in $\mathbb{R}^D$ as
\begin{align}
0 &= \partial_t \scF + \nabla\cdot [\vv_{\vx} \scF], \nonumber \\
  &= \partial_t \scF + \vv_{\vx} \cdot \nabla \scF + \scF \nabla\cdot \vv_{\vx}. \label{eq:Rdcontinuityexpansion}
\end{align}
We express $\nabla \scF$ as 
\begin{equation}
\nabla \scF = \frac{1}{r}\nabla_{\scS} \scF + \hr \frac{\partial \scF}{\partial r},
\end{equation}
where $\nabla_{\scS} \scF$ is the component of the gradient of $\scF$ along the surface $\scS$, as has been described in the main text. Since $\vv_{\vx} \cdot \hr = 0$, and $\nabla\cdot \uu{W}\vx=0$, we can simplify Eq. (\ref{eq:Rdcontinuityexpansion}) to
\begin{dmath}\label{eq:derivationintermediate}
\partial_t \scF + (1/r)\{\uu{W}\vx/r +   [\vrho - \vx (\vrho \cdot \vx)/r^2]\} \cdot \nabla_{\scS} \scF \\ +  \scF \nabla \cdot [\vrho - \vx (\vrho \cdot \vx)/r^2] = 0.
\end{dmath}
Now,
\begin{align*}
\nabla \cdot \left[\vrho - \frac{\vx (\vrho \cdot \vx)}{r^2}\right] &= -\nabla \cdot \left(\frac{\vx (\vrho \cdot \vx)}{r^2}\right), \\
																												 &= - \left[\frac{\vrho \cdot \vx}{r^2} \nabla \cdot \vx + \vx \cdot \nabla \frac{\vrho \cdot \vx}{r^2} \right], \\
																												 &= -\frac{\vx \cdot \vrho}{r^2} (D-1).
\end{align*}
Also note that 
\begin{equation*}
[\vrho - \vx (\vrho \cdot \vx)/r^2] \cdot \nabla_{\scS} \scF = \vrho \cdot \nabla_{\scS} \scF,
\end{equation*}
since $\vx\cdot \nabla_{\scS} \scF =0$ by the definition of $\nabla_{\scS}\scF$. Thus, Eq. (\ref{eq:derivationintermediate}) simplifies to
\begin{equation}
\frac{\partial \scF}{\partial t} + \frac{1}{r}(\uu{W}\hat r + \vrho)\cdot \nabla_S\scF - (D-1)\scF \hat r \cdot \vrho.
\end{equation}
Integrating the above equation over $r$ from $1-\epsilon$ to $1+\epsilon$ for small $\epsilon$, gives the desired result Eq. (\ref{eq:distcontinuity}).

\section{Critical coupling constant for even dimensions}\label{apx:evenDproof}

We now determine $K_c$ for even $D=2\Lambda$ as that value of $K$ such that $|\vrho| \to 0$ with $|\vrho|\neq 0$ as $K\to K_c$ from above. For notational simplicity, we write $\rho=|\vrho|$.
As discussed earlier in Sec. \ref{sec:evenD}, $\uu{W}$ can be written as $\uu{W}=\uu{R}^T\uu{D}\uu{R}$, where $\uu{R}$ is an orthogonal matrix, and $\uu{D}$ is a block-diagonal matrix with the $j^{\text{th}}$ block being a $2\times 2$ antisymmetric matrix with nonzero entries $\om_j$ and $-\om_j$ for all $j\in\{1,\hdots,\Lambda\}$. By construction, we choose $G(\uu{W})$ to be a distribution invariant to rotation, and hence we can rewrite $G(\uu{W})$ as
\begin{equation}
G(\uu{W})=g(\{\om_i\})U[\uu{R}],
\end{equation}
where $\{\om_i\}=\{\om_1,\om_2,\hdots\om_\Lambda\}$ represents the set of associated frequencies for each of the $2\times2$ blocks of $\uu{D}$, with $g(\{\om_i\})$ representing the joint distribution of these frequencies, and $U[\uu{R}]$ representing the uniform distribution of orthogonal matrices (corresponding to the Haar measure on the group of orthogonal matrices).
We then write Eq. (\ref{eq:rhosqoriginal}) as, 
\begin{equation}\label{eq:rhosqfreq}
\rho^2 = \int_{\uu{R}} \int_{\Gamma} \gamma(\{\om_i\},\{\rho_i\}) g(\{\om_i\})d\om_1\hdots d\om_\Lambda U[\uu{R}]d\uu{R}.
\end{equation}
Recall that $\rho_k^2$ is the sum of the squares of the magnitudes of component $2k-1$ and component $2k$ of $\vrho$ in the basis that block-diagonalized $\uu{W}$, corresponding to the components of $\vrho$ that are acted on by the $k^{\text{th}}$ block of $\uu{W}$.

\begin{figure}
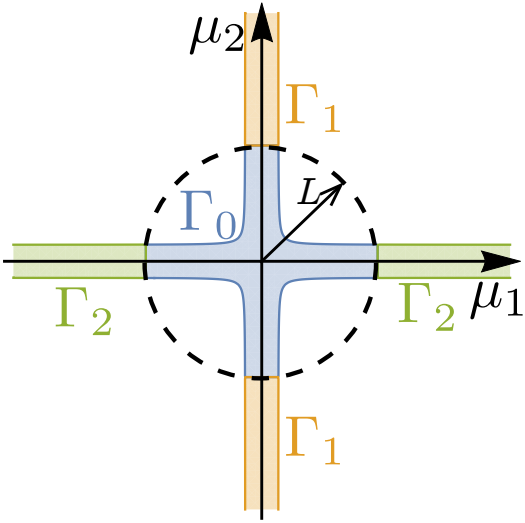

\includegraphics[width=.8\columnwidth]{{{domain_plot_new}}}
\caption{The shaded regions (in blue, green and orange) correspond to the domain $\Gamma$ in which $\sum_k{1/\mu_k^2} >1$ for the case of $D=4$ ($\Lambda=2$) in the $\{\mu_1,\mu_2\}$-space. The subdomain $\Gamma_0$, shown in blue, is the part of $\Gamma$ inside the circle of radius $L$; and the subdomains $\Gamma_i$ are the parts of the domain $\Gamma$ that lie outside $\Gamma_0$ which do not contain the $\mu_i$ axis ($\Gamma_1$ is shown in orange, and $\Gamma_2$ in green). The width of the strips in $\Gamma$ far away from the origin is $1$, hence the volume of the subdomain $\Gamma_0$ will scale as $\scO(L^{\Lambda-1})$ for large $L$. }
\label{fig:domain}
\end{figure}

Define $\mu_i=\om_i/(K\rho_i)$. In $\{\mu_i\}$-space, $\Gamma$ is the region $\sum_k{1/\mu_k^2} >1$, shown in Fig. \ref{fig:domain}.

Then
\begin{equation}\label{eq:rhosqmus}
\rho^2 = \int_{\uu{R}} \int_{\Gamma} \gamma g(\{\mu_i K \rho_i\}) K^\Lambda \rho_1 d\mu_1\hdots \rho_\Lambda d\mu_\Lambda U[\uu{R}]d\uu{R}.
\end{equation}

We next define a quantity $L\gg 1$ by choosing $L\sim\scO(\rho^{-1/2})$. Since we are interested in $\rho\to 0$, $L\to\infty$. Taking motivation from the shape of the domain $\Gamma$ shown in Fig. (\ref{fig:domain}), we express $\Gamma$ as the disjoint union of $\Gamma_0, \Gamma_1, \hdots, \Gamma_\Lambda$, where $\Gamma_0$ is the component of $\Gamma$ within the dashed circle of radius $L$ in Fig. \ref{fig:domain}, and for $j\geq 1$, $\Gamma_j$ is the region for which $|\mu_j| \lesssim 1$ and $|\mu_k|\geq L$ for all $k\neq j$. 

Note that the left-hand side of Eq. (\ref{eq:rhosqoriginal}) is $\rho^2$, hence we can ignore terms on the right-hand side of order smaller than $\scO(\rho^2)$. We now show that the contribution from $\Gamma_0$ is of a smaller order than this.
By construction, in the subdomain $\Gamma_0$, $|\mu_i|\leq L$, and hence $\mu_i K \rho_i \sim \scO(\sqrt{\rho}) \to 0$ as $\rho\to 0$. Further, $\gamma=\vrho\cdot\vx^F\leq \rho$. Thus the contribution $\mathcal{I}_{\Gamma_0}$ to the integral in Eq. (\ref{eq:rhosqmus}) from the subdomain $\Gamma_0$ will be 
\begin{align*}
\mathcal{I}_{\Gamma_0} &\lesssim \int_{\Gamma_0} \rho g(0,\hdots,0) K^\Lambda \rho_1 d\mu_1\hdots \rho_\Lambda d\mu_\Lambda \\
                           &\sim \scO[\rho^{\Lambda+1}\text{Volume}(\Gamma_0)].
\end{align*}
Since $L\gg 1$, the volume of $\Gamma_0$ will scale as $\scO(L^{\Lambda-1})\sim\scO(\rho^{-(\Lambda-1)/2})$. Thus $\mathcal{I}_{\Gamma_0} \sim \scO(\rho^{(\Lambda+3)/2})$, which for $D>2$ is negligible compared to $\rho^2$ and the contributions to the integrals in Eq. (\ref{eq:rhosqmus}) from the subdomains $\Gamma_j$.
Since $K_c$ for $D=2$ is already known (e.g. Ref. \cite{Strogatz2000}), we focus on the cases $D\geq 4$, and hence will ignore the contribution from the subdomain $\Gamma_0$. By symmetry, each $\Gamma_i$ will give the same contribution. Hence, without loss of generality, we will look at the contribution from the subdomain $\Gamma_1$, and will append a factor of $\Lambda$. We will also only look at $\mu_i>0$ and will hence append a factor of $2^\Lambda$.
\begin{align*}
\rho^2 = \Lambda 2^{\Lambda}\int_{\uu{R}} \int_{\mu_1=0}^1 \int_{\mu_2=L}^\infty \hdots \int_{\mu_\Lambda=L}^\infty &\gamma  g(\{\mu_i K \rho_i\}) K^\Lambda \\
																																																&\rho_1 d\mu_1\hdots \rho_\Lambda d\mu_\Lambda U[\uu{R}]d\uu{R}
\end{align*}

Going back to Eq. (\ref{eq:Hgamma}), we rewrite it as
\begin{equation}
\sum_k{\frac{1}{\gamma^2/\rho_k^2 + \mu_k^2}} = 1.
\end{equation}
In the subdomain $\Gamma_1$, $\mu_i \gg \mu_1$ for all $i\geq 2$, and thus we can use the above equation in the small $\rho_1$ approximation, 
\begin{align}
1 &\cong \frac{1}{\gamma^2/\rho_1^2 + \mu_1^2}, \\ 
\gamma &\cong \sqrt{\rho_1^2 (1-\mu_1^2).}
\end{align}
Also, $\mu_1<1$ implies $\mu_1 K \rho_1 \sim\scO(\rho)\to 0$ as $\rho\to 0$. Thus,
\begin{dmath}
\rho^2 = \Lambda 2^{\Lambda}K^\Lambda \int_{\uu{R}} \int_{\mu_2=L}^\infty \hdots \int_{\mu_\Lambda=L}^\infty \int_{\mu_1=0}^1  \rho_1\sqrt{1-\mu_1^2} d\mu_1  \\ g(0,\{\mu_i K \rho_i\}_{i=2}^\Lambda) \rho_1\hdots\rho_\Lambda d\mu_2\hdots d\mu_\Lambda U[\uu{R}]d\uu{R}.
\end{dmath}
We then change variables back to $\om_i$ for each of the $\mu_i$ integrals for $i=2\hdots\Lambda$, and explicitly evaluate the integral over $\mu_1$. The lower limits of the integrals change from $L$ to $LK\rho_i$ which goes to zero in the limit of small $\rho$, since $L\sim\scO(\rho^{-1/2})$. Thus

\begin{alignat}{3}
&\rho^2 &&= \Lambda 2^{\Lambda} K \frac{\pi}{4}\int_{\uu{R}} \int_{0}^\infty \hdots \int_{0}^\infty   && g(0,\{\om_i\}_{i=2}^\Lambda)  \rho_1^2  \nonumber \\
&				&&																																													&&d\om_2\hdots d\om_\Lambda U[\uu{R}]d\uu{R}, \nonumber \\
&       &&= \Lambda 2^{\Lambda} K \frac{\pi}{4}\int_{\uu{R}}  \rho_1^2  \frac{\tilde g(0)}{2^{\Lambda-1}} U[\uu{R}]&&d\uu{R}, \label{eq:rhosqsimplification}
\end{alignat}
where 
\begin{equation}\label{eq:gtilde0positivedefinition}
\tilde g(0) = 2^{\Lambda-1} \int_{0}^\infty \hdots \int_{0}^\infty   g(0,\{\om_i\}_{i=2}^\Lambda)  d\om_2\hdots d\om_\Lambda,
\end{equation}
equivalent to the definition given earlier in Eq. (\ref{eq:gtilde0}).
Since $U[\uu{R}]$ is the uniform distribution, thus by symmetry
\begin{align}
\Lambda \int_{\uu{R}}  \rho_1^2 U[\uu{R}]d\uu{R} &= \int_{\uu{R}}  \sum_{k}^\Lambda \rho_k^2 U[\uu{R}]d\uu{R}  \nonumber \\
                                                 &= \int_{\uu{R}}  \rho^2 U[\uu{R}]d\uu{R}  \nonumber \\
																								 &= \rho^2 \label{eq:rho1integral}
\end{align}
Inserting Eqs. (\ref{eq:gtilde0positivedefinition}) and (\ref{eq:rho1integral}) into Eq. (\ref{eq:rhosqsimplification}) gives us
\begin{equation*}
\rho^2 = 2^{\Lambda} K_c \frac{\pi}{4} \rho^2 \frac{\tilde g(0)}{2^{\Lambda-1}}.
\end{equation*}
Since we are in the limit of small but \emph{nonzero} $\rho$, we can cancel $\rho^2$ from both sides to obtain the desired result in Eq. (\ref{eq:kcgtilde})

\bibliography{Kuramoto_paper}

\end{document}